\documentclass[aps,amsmath,amssymb,showkeys]{revtex4}
\usepackage[dvips]{graphicx,color}
\usepackage{times}
\usepackage{mathrsfs}
\usepackage{amsmath}
\usepackage{amssymb}
\usepackage{graphicx}
\usepackage{epsfig}
\usepackage{dcolumn}
\usepackage{bm}
\usepackage{xcolor}
\usepackage{hyperref}
\hypersetup{colorlinks=true,linkcolor=red,citecolor=blue, urlcolor=teal}
\newcommand{\orcid}[1]{\href{https://orcid.org/#1}{\includegraphics[width=8pt]{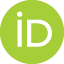}}}

\begin{document}	
\title{Galactic rotation curves of spiral galaxies and dark matter in 
$f(\mathcal{R},T)$ gravity theory}	

\author{Gayatri Mohan \orcid{0009-0008-0654-5227}}
\email[Email: ]{gayatrimohan704@gmail.com}

\author{Umananda Dev Goswami \orcid{0000-0003-0012-7549}}
\email[Email: ]{umananda2@gmail.com}	

\affiliation{Department of Physics, Dibrugarh University, Dibrugarh 786004, 
Assam, India}	

\begin{abstract}
Galactic rotation curve is a powerful indicator of the state of the 
gravitational field within a galaxy. The flatness of these curves indicates 
the presence of dark matter in galaxies and their clusters. In this paper, we 
focus on the possibility of explaining the rotation curves of spiral galaxies 
without postulating the existence of dark matter in the framework of 
$f(\mathcal{R},T)$ gravity, where the gravitational Lagrangian is written by 
an arbitrary function of $\mathcal{R}$, the Ricci scalar and of $T$, the trace 
of energy-momentum tensor $T_{\mu\nu}$. We derive the gravitational field 
equations in this gravity theory for the static spherically symmetric 
spacetime and solve the equations for metric coefficients using a specific 
model that has minimal coupling between matter and geometry. The orbital 
motion of a massive test particle moving in a stable circular orbit is 
considered and the behavior of its tangential velocity with the help of the 
considered model is studied. We compare the theoretical result predicted by 
the model with observations of a sample of nineteen galaxies by generating 
and fitting rotation curves for the test particle to check the viability of 
the model. It is observed that the model could almost successfully explain 
the galactic dynamics of these galaxies without the need of dark matter at 
large distances from the galactic center.
\end{abstract}



\maketitle

\section{Introduction}
\label{sec1}
The significant mismatch between the observed and dynamically estimated 
masses of different astrophysical systems such as galaxies and clusters of 
galaxies raises the issue of dark datter (DM), which is one of the leading 
theoretical difficulties in modern astrophysics. The pioneering works and 
observational data show that the total mass density of the universe is not 
dominated by thebaryonic matter, instead by some unseen, unusual form of 
matter called DM, which is needed for the explanation of both the galactic 
dynamics as well as the large scale structures in the universe \cite{Mar}. 
The presence of such mysterious matter was first pointed out by J.~H.~Oort in 
the early 1930s \cite{J_3}. He studied the doppler shifts of some stars of 
our galaxy moving near the galactic plane and thus by computing their 
velocities he argued for the presence of more mass inside the galaxy than the 
observed mass to hold the stars in their respective orbits. In 1933, a Swiss 
astronomer F.\ Zwicky \cite{F_5a} suggested similar hidden matter from his 
study on the Coma cluster of galaxies. By employing virial theorem he 
estimated the virial mass (total mass) of the cluster from the velocity 
dispersions of galaxies using doppler shift and then comparing the virial 
mass of the cluster to the luminous mass calculated from $M/L$ ratio, he 
found an inequality of these two masses which implies a large mass discrepancy 
in the cluster \cite{Gar_4,F_5a,F_1, F_5}. In 1939, H.\ Babcock \cite{H_9} 
studied the rotation curve of M31 about 20 kpc away from the center of the 
galaxy and by estimating mass distribution of it, he proposed for a large 
amount of mass in the exterior part of the galaxy. Outstanding investigations 
on the mass distributions in galaxies were published during late 1960s to 
early 1980s through the observations of rotational velocities of stars in 
some galaxies by M.\ Roberts \cite{M_10}, K.\ Freeman \cite{K_12}, 
J.~C.~Jackson \cite{J_8}, Roberts and Rots \cite{M_10a}, Rubin and Ford 
\cite{ V_11, V_11a} etc. They noted that the mass of a galaxy must be larger 
than its detectable mass and thus they confirmed about an undetectable, 
significant, additional mass density at large distances from the center. From 
the analyses of rotation curves based on radio observatory data, D.~H.~Rogstad 
and G.~Shostak \cite{R_D} also suggested the necessity of low luminosity 
material in the outer regions of M33, M101, NGC 6946, NGC 2403 and IC 342 
galaxies.
	
As already clear from above, two distinguished observational findings, the 
mass discrepancies in clusters of galaxies and the flat shapes of rotation 
curves of spiral galaxies are two striking evidences for the existence of DM 
within these structures \cite{L_2,Anna, C15,M_5b}. Moreover, another 
observational discovery within General Relativity (GR) is the gravitational 
lensing, in which distorted image of a distant light source is formed due to 
the deflection of light by a massive gravitating system such as a cluster of 
galaxies, is also a strong evidence of DM within the lensing cluster. It is 
also a powerful tool to determine the amount of mass within the lensing 
cluster \cite{Gar_4,Ca}. Further, rotation curves, the plots of tangential 
velocities of stars and gas clouds rotating around the center of galaxies as 
a function of the distance from the center, are directly related to the 
overall mass distributions of galaxies and are potent means to analyze the 
gravitational field within galaxies. These curves show that the tangential 
velocity of a star increases linearly with the distance from the center of the 
galaxy and attains approximately a constant value $v \approx$ $200$ -- $300$ 
km/s \cite{L_2,THar} up to $5$ -- $6$ times \cite{Mar} of the luminous radius 
of a galaxy. It infers that a visible galaxy is embedded in an extended 
spherical matter distribution called the galactic halo (DM halo), where due 
to a constant $v$, the mass $M(r)$ within the radius $r$ increases linearly 
with $r$ according to the relation $M(r)= v^2 r/G$ and it causes  density 
variation as $1/r^2$ for a spherically symmetric model \cite{Tg,V_11b,V_11c}. 
This variation is exceptional from the Newtonian gravity, which yields 
Keplerian decrease in velocity at a large distance from the center. 
	
Most importantly, rotation curves hint the failure of GR at large distances, 
as to fit the flat parts of the curves GR needs DM in the outer parts of the 
galaxies. But till now no experimental evidence has been found or no direct 
evidence of DM has been detected. Consequently, attempts have been made for 
searching the alternatives of Newtonian Gravity and GR. As a result, based on 
the modification of Newtonian gravity or of GR different theoretical models 
have been proposed to understand observations both at cosmological and 
astrophysical scales \cite{C15,Mil,Ad,RH}. M.~Milgrom in $1983$, proposed the 
Modified Newtonian Dynamics (MOND) \cite{Mil} as a possible means to describe 
the hidden mass in galaxies and their clusters. Basis of this modification is 
that the acceleration $a$ of a particle follows the relation 
$a^2/a_0 \approx MG/r^2$ within the limit of small acceleration $a\ll a_0$ at 
a distance $r$ from a mass $M$, where $a_0$ is a constant with the dimension 
of acceleration. Similarly, the theory of GR has been generalised by modifying 
the geometric part of standard Einstein-Hilbert (E-H) action with the higher 
order curvature invariants \cite{SC} to understand both cosmological and 
astrophysical observations. Among these modified theories, termed as Modified 
Theories of Gravity (MTGs), the simplest one is the $f(\mathcal{R})$ gravity, 
introduced by H.~A.~Buchdahl in $1970$ \cite{HB}. He considered a modification 
in the Lagrangian $\mathcal{L}$ of the E-H action and formulated the theory 
by choosing a nonlinear function of the Ricci scalar $\mathcal{R}$ in place of 
$\mathcal{R}$ and discussed the singularity problem of the Robertson-Walker 
cosmological model. Subsequently, several $f(\mathcal{R})$ gravity models have 
been developed with different $f(\mathcal{R})$ functions, which are playing a 
significant role in successful explanation of cosmic as well as galactic 
dynamics, large scale structures without any exotic component in the universe.
Particularly, for example, the galactic dynamic of a test particle without DM 
has been studied in Refs.~\cite{C15,Bra,Ra6,Cap,Y11,cf,SSh,MkM,Ugu, np, jr}.
Moreover, many investigations have also been conducted by analysing the 
scalaron's property in $f(\mathcal{R})$ gravity theory in many literatures 
(see e.g.~\cite{bk,JF,TT,Sca}) to explain DM as the scalar field manifestation 
of spacetime curvature.  
	
The theory of GR has also been modified or extended by amending both the 
geometry and mass-energy content of the Lagrangian $\mathcal{L}$ of the E-H 
action. One of such interesting extended theory of gravity is the 
$f(\mathcal{R},T)$ gravity theory, proposed by T.~Harko et al.\ \cite{TH} in 
$2011$, where the gravitational Lagrangian $\mathcal{L}$ is a function of 
both the Ricci scalar $\mathcal{R}$ and the trace $T$ of the energy-momentum 
tensor. Thus this theory is based on the idea of possible coupling between 
the geometry and matter. The notable point of the theory is that 
$f(\mathcal{R},T)$ gravity models depend on the source term and source term 
has a general expression in terms of matter Lagrangian $\mathcal{L}_m$. 
Therefore, there would be a particular set of field equations for every choice 
of matter Lagrangian. The gravitational field equations in the theory contain 
two fluid sources: one is the pure or perfect fluid source which is assumed as 
the matter source and other is due to the interaction between matter and 
geometry, called the exotic fluid source. Harko et al.~have demonstrated that 
the covariant derivative of energy-momentum tensor does not vanish in the 
theory giving the non-geodesic motion of a massive test particle and also it 
has been noted that an extra force remains effective because of the 
matter-geometry coupling developed in the theory. Later, Moraes presented that 
the extra force may vanish for $\mathcal{L}_m = -\,p $ in the non-relativistic 
dust matter case \cite{pm}. Under analogous noninteracting two fluid 
structures as presented by Harko et al., S.~Chakraborty has described 
\cite{S7} that the conservation of energy-momentum tensor can be taken into 
account for some specific forms of the $f(\mathcal{R},T)$ function. Hence for 
such cases the geodesic motion of a test particle in $f(\mathcal{R},T)$ 
gravity can be achieved. The $f(\mathcal{R},T)$ theory encourages 
investigations in different issues of cosmology \cite{MJ,Hs,Eb,Mjs} as well as 
of astrophysics \cite{S.K.,Jmz,Ms} including the DM in small \cite{AVZ,Ps,MPS} 
and large scale structures \cite{Bl,TKO}. Maurya et al.~have performed a study 
on compact structures in $f(\mathcal{R},T)$ gravity with anisotropic matter 
distribution and found it as a justified theory in explaining such 
astrophysical objects \cite{S.K.}. Galactic rotation curves with non-minimal 
coupling between matter and geometry have been investigated by 
T.\ Harko~\cite{THar} and showed the dependence of tangential velocity upon 
coupling function between matter and geometry. Further, H.\ Shabani and 
M.\ Farhou~\cite{H13} have studied the $f(\mathcal{R},T)$ cosmological models 
in phase space and analyzed for the minimal, non-minimal and purely 
non-minimal theories giving the conclusion that the non–minimal theory can 
have acceptable cosmological solutions. Effects of DM on galactic scale have 
been suitably elaborated by R.~Zaregonbadi and his collaborators considering 
a $f(\mathcal{R},T)$ model of minimal matter and geometry coupling \cite{Ra6}.
They showed that the interaction mass obtained from the modification of 
gravity with minimal model leads to constant tangential velocity in the halo 
of a galaxy and extended their work to study light-deflection angle also. 
H.~Shabani and P.~H.~R.~S.~Moraes \cite{H3} investigated the dark matter issue 
from the $f(\mathcal{R},T)$ modified gravity by considering both the minimal 
coupling and the non-minimal coupling models, and argued the possibility to 
attain flat galaxy rotation curves only from the non-minimal models. Such 
remarkable studies have motivated us to understand DM through galactic 
rotation curves derived from coupling of matter and geometry. The aim of the 
study is to investigate galactic dynamics with a minimally coupled 
$f(\mathcal{R},T)$ model and test the result predicted by the model with 
observations by fitting it with observational rotation velocity data of a 
sample of some high surface brightness (HSB) and low surface brightness
(LSB) galaxies.   
	
In our present work, as we intend to study the galactic rotation curves in 
the framework of $f(\mathcal{R},T)$ gravity theory and hence to explain the 
mystery of DM as a large-scale modification of gravity, we consider a test 
particle moving around a galaxy in a stable circular orbit. Here we employ the 
geodesic equations to obtain the rotation curves of galaxies in spherically 
symmetric spacetime and derive rotational velocity with a $f(\mathcal{R},T)$ 
model that has minimal coupling between matter and geometry. We organize our 
rest of the paper as follows. In Section \ref{sec2}, we derive the field 
equations in the $f(\mathcal{R},T)$ gravity and then transform these equations 
in the form of modified Einstein equations. In Section \ref{sec3}, the 
solutions of the modified field equations for the static spherically symmetric 
metric is presented. In Section \ref{sec4}, we considered a minimally coupled 
$f(\mathcal{R},T)$ model and the metric coefficients are expressed in terms of 
it. The orbital motion of a test particle is analysed in Section \ref{sec5}. 
Finally we conclude our work in Section \ref{sec6}. In this work we use the 
unit $c = 1$ and the metric signature $(-,+,+,+)$. 
	
\section{Modified Einstein field equations in $f(\mathcal{R},T)$ gravity}
\label{sec2}
The action in the $f(\mathcal{R},T)$ gravity theory takes the form:
\begin{equation} 
S = \frac{1}{2\kappa^2}\int\!\sqrt{-\, g}\,{f(\mathcal{R},T)}\, d ^4x + S_m,
\label{eq1}
\end{equation} 
where $S_m = \int\!\sqrt{- \, g}\, \mathcal{L}_m\, d^4x$ is the matter part of 
the action with $\mathcal{L}_{m}$ as the matter Lagrangian density of matter 
field, $g$ is determinant of the metric $g_{\mu\nu}$ and 
$\kappa^2 = 8 \pi G = 1/M^2_{pl}$. $G$ and $M_{pl}$ are the gravitational 
constant and Planck mass (reduced) respectively. The energy-momentum tensor 
of matter field can be defined as 
\begin{equation} 	
T_{\mu\nu} = -\,\frac{2\, \delta(\sqrt{- \, g}\mathcal{L}_{m})}{\sqrt{- \, g}\ \delta g^{\mu\nu}} = -\, 2\ \frac{\partial \mathcal{L}_{m}}{\partial g^{\mu\nu}} + g_{\mu\nu}\, \mathcal{L}_{m}.
\label{eq2}
\end{equation}     
From this equation we obtain the trace of $T_{\mu\nu}$ as
\begin{equation}
T  =  -\, 2\, g^{\mu\nu}\frac{\partial \mathcal{L}_{m}}{\partial g^{\mu\nu}} + 4\,\mathcal{L}_m.
\label{eq3}
\end{equation}
In this case we assumed that the matter Lagrangian density $\mathcal{L}_m$ 
does not depend on the derivative of the metric tensor $g_{\mu\nu}$, instead 
it depends on $g_{\mu\nu}$ only. Now, the variation of action \eqref{eq1} 
with respect to $g_{\mu\nu}$ results in the field equations of 
$f(\mathcal{R},T)$ gravity as 
\begin{equation}
F_R\, \frac{\delta{R}}{\delta{g^{\mu\nu}}} - \frac{1}{2}\,f(\mathcal{R},T)\, 
g^{\mu\nu} = \kappa^2 T_{\mu\nu} - F_T\,\frac{\delta{T}}{\delta{g^{\mu\nu}}},
\label{eq4}
\end{equation}
where $$F_R = \frac{\partial f(\mathcal{R},T)}{\partial R},\;\; 
F_T = \frac{\partial f(\mathcal{R},T)}{\partial T},$$ 
\begin{equation}
\frac{\delta R}{\delta g^{\mu\nu}} = R_{\mu\nu} - \nabla_\mu\nabla_\nu + g_{\mu\nu}\square\;\; \text{and}\;\; 
\frac{\delta T}{\delta g^{\mu\nu}} = \frac{\delta(g^{\alpha\beta} T_{\alpha\beta})}{\delta g^{\mu\nu}} = T_{\mu\nu} + \theta_{\mu\nu}. 
\label{eq5}
\end{equation}
Here, $R_{\mu\nu}$ is the Ricci tensor, $\nabla_{\mu}$ is the covariant 
derivative related to Levi-Civita connection $\Gamma$, 
$\square = \nabla_{\mu} \nabla^ {\mu}$ represents the d'Alembertian operator 
and   
\begin{equation}
\theta_{\mu\nu}= g^{ab}\,\frac{\delta{T}_{ab}}{\delta{g^{\mu\nu}}}= -\,2\,T_{\mu\nu} + g_{\mu\nu}\, \mathcal{L}_{m} - 2\,g^{ab}\frac{\partial^2{\mathcal{L}_m}}{\partial{g^{\mu\nu}}\partial{g^{ab}}}
\label{eq6}
\end{equation}
is the energy tensor~\cite{TH,FG}. The use of equation \eqref{eq5} reduces the 
equation \eqref{eq4} to the following form: 
\begin{equation}
F_R\, R_{\mu\nu} - \frac{1}{2}\, f(\mathcal{R},T)\, g_{\mu\nu} + \left(g_{\mu\nu} \square - \nabla_\mu \nabla_ \nu\right)F_R = \kappa^2 T_{\mu\nu} -  F_T\left(T_{\mu\nu} + \theta_{\mu\nu}\right).
\label{eq7}	
\end{equation}
As already mentioned, the basis of $f(\mathcal{R},T)$ theory is the coupling 
between matter and geometry, which shows by equations \eqref{eq7} that the 
field equations of the theory depend on energy tensor $\theta_{\mu\nu}$. Thus, 
corresponding to different choice of $\mathcal{L}_m$, a different set of field 
equations would be generated \cite{M3,JB}. $\mathcal{L}_m$ can be written in 
terms of energy density $\rho$ or in terms of thermodynamic pressure $p$ 
\cite{Fra}. The choice of the matter Lagrangian $\mathcal{L}_m$ equal to $p$ 
results $\theta_{\mu\nu} = -\,2\, T_{\mu\nu} + p\, g_{\mu\nu}.$ Hence, from 
equation \eqref{eq7} by assuming the energy-momentum tensor $T_{\mu\nu}$ for 
a dustlike matter field, we finally obtain the field equations of 
$f(\mathcal{R},T)$ theory as follows \cite{R5,Ra6}: 	
\begin{equation}
F_R\, R_{\mu\nu} - \frac{1}{2}\, f(\mathcal{R},T)\, g_{\mu\nu} + \left(g_{\mu\nu} \square - \nabla_\mu \nabla_ \nu\right)F_R = \left(\kappa^2  + F_T\right)T_{\mu\nu}.
\label{eq8}	
\end{equation}
With a simple algebric manipulation, these field equations \eqref{eq7} 
can be written as the modified Einstein equations with an effective 
energy-momentum tensor $T_{\mu\nu}^{eff}$ as
\begin{equation} 
G_{\mu\nu} = 8\pi\, G_{eff}\, T_{\mu\nu}^{eff},
\label{eq9}
\end{equation}
where $ G_{eff} = G/F_R$ and $T_{\mu\nu}^{eff} = T_{\mu\nu} + T_{\mu\nu}^I. 
~T_{\mu\nu}^I$ represents the matter-curvature interaction term defined as 
interaction energy-momentum tensor and may act as a new matter source induced 
by the $f(\mathcal{R},T)$ action that behaves as an extra effective fluid of 
purely geometric origin \cite{R5,S7}. It is given as 
\begin{equation}
T_{\mu\nu}^I =\ \frac{1}{\kappa^2}\left[F_T\, T_{\mu\nu} + \frac{1}{2}\left(f(\mathcal{R},T) - F_R R\right)g_{\mu\nu} + \left(\nabla_\mu\nabla_\nu - g_{\mu\nu} \square\right)F_R\right].
\label{eq10} 	
\end{equation}
	
Refs.~\cite{TH,JB,R5} have discussed the conservation of energy-momentum 
tensor and its consequences on the motion of a test particle in 
$f(\mathcal{R},T)$ gravity. Ref.\ \cite{TH} shows that though an additional 
force perpendicular to four velocity appears in the equation of motion of a 
test particle in $f(\mathcal{R},T)$ theory of gravity resulting the motion of 
the particle as non-geodesic, in case of dust (pressureless fluid), similar 
to GR, the conservation of the energy-momentum tensor in the theory holds 
good. But it has been corrected in Ref.\ \cite{JB} that as the covariant 
divergence of energy-momentum tensor is not conserved even for pressureless 
fluid and hence it leads to non-geodesic path to the particle. In general, 
the conservation equation of the effective energy-momentum tensor 
$T_{\mu\nu}^{eff}$ for $f(\mathcal{R},T)$ theories discussed in 
Ref.\ \cite{R5} shows that by using the usual energy-momentum conservation 
$\nabla^\mu T_{\mu\nu}=0$ and $\nabla^\mu G_{\mu\nu}=0$ obtained from the 
Bianchi relation for Ricci scalar, the covariant divergence equation for the 
$T_{\mu\nu}^{eff}$ can be achieved in the following form~\cite{R5,H3,TK1}:  
\begin{equation}
\nabla^\mu  T_{\mu\nu}^{eff} = \frac{1}{\kappa^2}\left[ T_{\mu\nu}\nabla^\mu F_T + \frac{1}{2} F_T \nabla_\nu T +  G_{\mu\nu}\nabla^\mu F_R\right].  
\label{eq11} 
\end{equation}     
Under the consideration that 
$T_{\mu\nu}\nabla^\mu F_T + \frac{1}{2} F_T \nabla_\nu T = 0$, 
equation \eqref{eq11} reduces to
\begin{equation} 
\nabla^\mu  T_{\mu\nu}^{eff} = \frac{G_{\mu\nu}}{\kappa^2}\, \nabla^\mu F_R\,.
\label{eq12}
\end{equation}
This consideration imposes some restrictions on the choice of functionality 
of $f(T)$ of $f(\mathcal{R},T)$. Equation \eqref{eq12} indicates that 
conservation of interaction energy-momentum holds good only for the constant 
value of $F_{R}$ which evidently confirms the geodesic motion of the test 
particle for the same form of field equations.
	
It is to be noted that because of the extra terms appearing in the field 
equations, the $f(\mathcal{R},T)$ theory of gravity can be regarded as a two 
fluid model \cite{pm}. Further, defining $T_{\mu\nu}^I$ for an anisotropic 
geometric matter distribution, we can express it as \cite{Bra}
\begin{equation}    
T_{\mu}^{\nu I} = \text{diag}\left(-\rho^I,\ P_r^I,\ P_t^I,\ P_t^I\right),
\label{eq13} 
\end{equation}  
where $\rho^I$ represents interaction energy density, $P_r^I$ and $P_t^I$ are 
the radial and tangential components of interaction pressure respectively. 
Thus we obtain,
\begin{equation} 
\rho^{eff} = \rho + \rho^I, \;\;\; P_r^{eff} = P_r^I,\;\;\; P_t^{eff} =  P_t^I.
\label{eq14}
\end{equation}
Here, $\rho^{eff}$ is effective energy density. As we have considered 
$T_{\mu\nu}$ for the dustlike matter, so both radial and tangential components 
of the effective pressure are equal to their respective interaction terms.	

\section{Gravitational Field Equations for Static Spherically Symmetric Mass 
Disribution}
\label{sec3}
As mentioned in Section \ref{sec1}, since our study mainly concentrates on 
the explanation of behaviour of galactic rotation curves of spiral galaxies in 
galactic halo regions without considering the existence of DM in it, we need 
to have a suitable mass distribution for such galaxies. Simply such a galaxy 
can be considered to have spherically symmetric distribution of matter having 
smooth gravitational field \cite{J9}. So we shall consider a general static 
spherically symmetric metric of the form:
\begin{equation}
ds^2 = - e^{2\mu} dt^2 + e^{2\nu} dr^2 + r^2d\theta^2 + r^2 \sin^2\theta\,d\varphi^2, 	
\label{eq15}
\end{equation} 
where $\mu$ and $\nu$ are metric potentials and are functions of $r$ only. In 
our present case the metric coefficients $e^{2\mu}$ and $e^{2\nu}$ can be 
determined by solving the modified Einstein field equations \eqref{eq9} as 
follows. The nonzero components of Einstein tensor $G_{\mu\nu}$ for the metric 
\eqref{eq15} can be obtained as
\begin{align}
\label{eq16}
G_0^0\, & =\, -\,\frac{2\,e^{-2\nu}}{r}\, \nu' + \frac{e^{-2\nu}}{r^2} -  \frac{1}{r^2},\\[5pt]
\label{eq17}
G_1^1\, & =\, \frac{2\, e^{-2\nu}}{r}\, \mu'  + \frac{e^{-2\nu}}{r^2} - \frac{1}{r^2},\\[5pt]
\label{eq18}
G_2^2\, =\,  G_3^3\,& =\, e^{-2\nu}\left[\mu'' + (\mu')^2 - \mu' \nu'\right] +  \frac{e^{-2\nu}}{r}\, (\mu' - \nu').
\end{align}
Using these equations \eqref{eq16}, \eqref{eq17} and \eqref{eq18} together 
with equation \eqref{eq9} and \eqref{eq14} we obtain the effective field 
equations as given by
\begin{align}
\label{eq19}
\frac{2\,e^{-2\nu}}{r}\, \nu' & - \frac{e^{-2\nu}}{r^2} + \frac{1}{r^2} = 8\, \pi\, G_{eff}\left(\rho + \rho^I\right),\\[10pt]
\label{eq20}
\frac{2\, e^{-2\nu}}{r}\, \mu' & + \frac{e^{-2\nu}}{r^2}-\frac{1}{r^2} = 8\, \pi\, G_{eff}\, P_r^I,\\[10pt]
\label{eq21}
e^{-2\nu}\Big[\ \mu'' & + (\mu')^2 - \mu' \nu' \Big]+  \frac{e^{-2\nu}}{r}\, 
(\mu' - \nu') = 8\, \pi\, G_{eff}\, P_t^I.
\end{align}
The interaction energy density $\rho^I$ and components of anisotropic 
interaction pressure $P_r^I$ and $P_t^I$ can be obtained from equation 
\eqref{eq10}, which can be expressed as	
\begin{align}
\label{eq22}
\rho^I\,& =\, \frac{1}{8\,\pi\,G}\left[F_{T}\, \rho + \frac{1}{2}\left(R\, F_R - f(\mathcal{R},T)\right)  + \left(F_R'' - \nu'F_R' + \frac{2}{r}\, F_R'\right)\,e^{-2\nu}\right],\\[10pt]
\label{eq23}
P_r^I\,& =\, \frac{1}{8\, \pi\, G} \left[\frac{1}{2}\left(f(\mathcal{R},T) - R\, F_R\right) +  \left(\mu'\,F_R' - \frac{2}{r}\, F_R'\right)e^{-2\nu}\right],\\[10pt]
\label{eq24}
P_t^I\,& =\, \frac{1}{8\, \pi\, G} \left[\frac{1}{2}\left(f(\mathcal{R},T) - R\, F_R\right) - \left(F_R'' + \frac{F_R'}{r} + (\mu' - \nu')\, F_R'\right)e^{-2\nu}\right].
\end{align}
To solve equations \eqref{eq19}, \eqref{eq20} and \eqref{eq21} for the metric 
coefficients, we just sum equation \eqref{eq19} and \eqref{eq20} which gives 
a differential equation for the functions $ \mu$ and $\nu$ in the form:  
\begin{equation}
\frac{\mu' + \nu'}{r\, e^{2\nu}} = 4\,\pi\,G_{eff}\left[\rho+\rho^I + P_r^I\right].
\label{eq25}
\end{equation}	

At this point it needs to be mentioned that in GR we have $\mu' + \nu' = 0$ 
and hence $e^{2\mu} e^{2\nu} = 1$. But as we are working with the 
$f(\mathcal{R},T)$ modified theory of gravity, which is an extension of GR, 
so have obtained equation \eqref{eq25}. Thus in our case, we should have 
$e^{2\mu}e^{2\nu}$ different from unity. It can be argued that if 
$\mu' + \nu'$ is a well-behaved expression it should have a solution of the 
following form \cite{Y11}:
\begin{equation}
e^{2\mu} e^{2\nu} = X(r)\,.
\label{eq26}
\end{equation} 	
Here, function $X(r)$ should be slightly different from unity to remain it in 
the vicinity of GR, so that the $f(\mathcal{R},T)$ gravity behaves similarly 
to GR in the low curvature regimes. To differ $X(r)$ from the unity by a 
small amount we assume that \cite{Y11}
\begin{equation}
X(r) = \left(\frac{r}{h}\right)^\delta\!\!,
\label{eq27}
\end{equation}
where $\delta$ is a small dimensionless  parameter, $h$ is the scale length 
of the system. With this form of $X(r)$ equation \eqref{eq26} takes the form:
\begin{equation}
2(\mu + \nu) = \ln\left(\frac{r}{h}\right)^\delta\!\!.
\label{eq28} 
\end{equation}
From which a differential equation for the metric potentials is achieved as
\begin{equation}
\mu'\ + \nu' =\,\frac{\delta}{2\,r}\,.
\label{eq29}
\end{equation}
Also $$e^{2\mu} e^{2\nu} = e^{2\mu+2 \nu} \cong 1+2(\mu + \nu)\,.$$ Hence we 
may write the product $e^{2\mu} e^{2\nu}$ approximately as	
\begin{equation}
\label{eq30} 
e^{2\mu} e^{2\nu} = X(r) \cong 1 + \ln\left(\frac{r}{h} \right)^\delta\!\!.
\end{equation}
Thus the product $e^{2\mu} e^{2\nu}$ is slightly greater than unity as 
expected. Now, using equation \eqref{eq29} in equation \eqref{eq25} we can 
express the metric coefficient $e^{2\nu}$ in the present case as
\begin{equation}
e^{2\nu} = \delta\left[8\,\pi\,G_{eff}\,r^2 \left(\rho+\rho^I + P_r^I\right)\right]^{-1}\!\!\!\!\!\!\!.
\label{eq31}
\end{equation}
And hence the coefficient of $g_{00}$ component of the metric can be written as
\begin{equation}
e^{2\mu} = 8\,\pi\,G_{eff}\,\delta^{-1} h^{-\delta}r^{\delta+2} \left(\rho+\rho^I+P_r^I\right).
\label{eq32}
\end{equation}
Later these metric coefficients will be determined in their explicit forms 
for our considered models of the $f(\mathcal{R},T)$ gravity.

It is observed that most of the baryonic matter of a normal spiral galaxy 
lies in the flattened disk in the form of stars, dust and interstellar gas. 
The mass to luminosity ratio of such galaxies increases with increasing 
distance from the center up to the outer regions of the galaxies \cite{V_11c}.
We may assume a spherically symmetric model for baryonic matter distribution 
in which the normal matter (baryonic) density in halo has a power law density 
variation given as some power of the radial distance $r$ as \cite{J9,Fra}
\begin{equation}
\rho(r) = \rho_0 \left(\frac{1}{r}\right)^\lambda\!\!\!\!,
\label{eq33}
\end{equation} 
where $\rho_0$ and $\lambda$ are positive constants, $\rho_0$ can be set as 
unity without loss of generality. The baryonic mass distribution within $r$ is
\begin{equation} 
M(r) = 4 \pi\! \int_{0}^{r}\!{r^2 \rho(r)}\, dr = 4 \pi \rho_0 \, \frac{r^{3-\lambda}}{3-\lambda}\,.
\label{eq34}
\end{equation}
It is clear that the pattern of baryonic mass distribution within $r$ depends 
on the value of the parameter $\lambda$.  
 
\section{The Metric Coefficients from the Model of Study}
\label{sec4}	
One can perform a study on $f(\mathcal{R},T)$ modified theory by employing 
different matter-geometry coupling models. The functional $f(\mathcal{R},T)$ 
in the theory may be modelled through minimally, non-minimally or purely 
non-minimally coupling of the matter and geometry sectors. In the minimally 
coupling case the functional form of $f(\mathcal{R},T)$ is assumed in the 
form of $f(\mathcal{R},T) = q(\mathcal{R}) + s(T)$, where $q(\mathcal{R})$ 
and $s(T)$ are arbitrary functions only of $\mathcal{R}$ and $T$ respectively. 
Whereas in the non-minimal and purely non-minimal cases $f(\mathcal{R},T)$ is 
assumed to have forms: $f(\mathcal{R},T) = q(\mathcal{R})[1 + s(T)]$ and 
$ f(\mathcal{R},T) = q(\mathcal{R})\, s(T)$ respectively \cite{Hs,R5,H13}. In 
our work, we consider that the matter sector and geometry sector in 
$f(\mathcal{R},T)$ gravity are coupled minimally and for the fulfilment of 
our desired result we consider the following minimally coupled model of 
$f(\mathcal{R},T)$ gravity: 
\begin{equation}
f(\mathcal{R},T) = a\,\mathcal{R}-b\,(-T)^{\beta},
\label{eq35}	
\end{equation}
where $a$ and $b$ are coupling constants that parameterize the departure of 
the model from GR and the constant parameter $\beta$ determines the strength 
of matter effect. For the existence of the model, parameters $a$, $b$ and 
$\beta$ should not be zero. In vacuum, beyond the galactic halo, where 
$T_{\mu\nu} = 0$, the model reduces to GR. Further, the model is respecting 
the conservation law also as discussed in Ref.~\cite{TH,R5}. For this model 
we have,
\begin{equation}
F_R =  a \;\;\; \text{and} \;\;\; F_T = b\,\beta\, \rho^{\beta - 1}.
\label{eq36}
\end{equation}
Thus using equation \eqref{eq22} the interaction energy density for this 
model \eqref{eq35} is obtained as
\begin{equation}
\rho^I = \frac{b\,\rho^{\beta}}{8\,\pi\, G} \left[\beta + \frac{1}{2}\right],
\label{eq37}
\end{equation}	

\begin{figure}[!h]
\centerline{
\includegraphics[scale = 0.35]{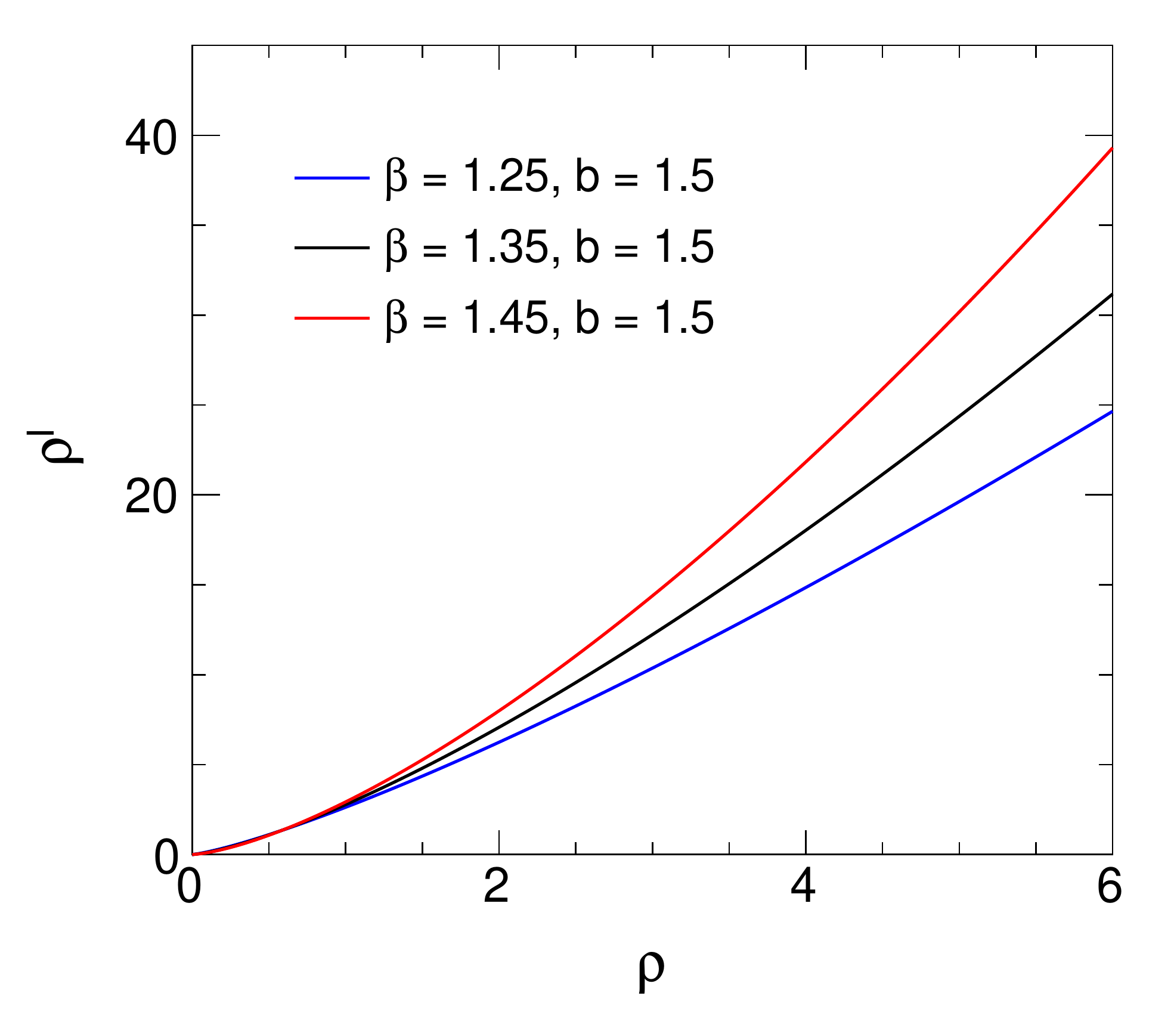}\hspace{0.50cm}
\includegraphics[scale = 0.35]{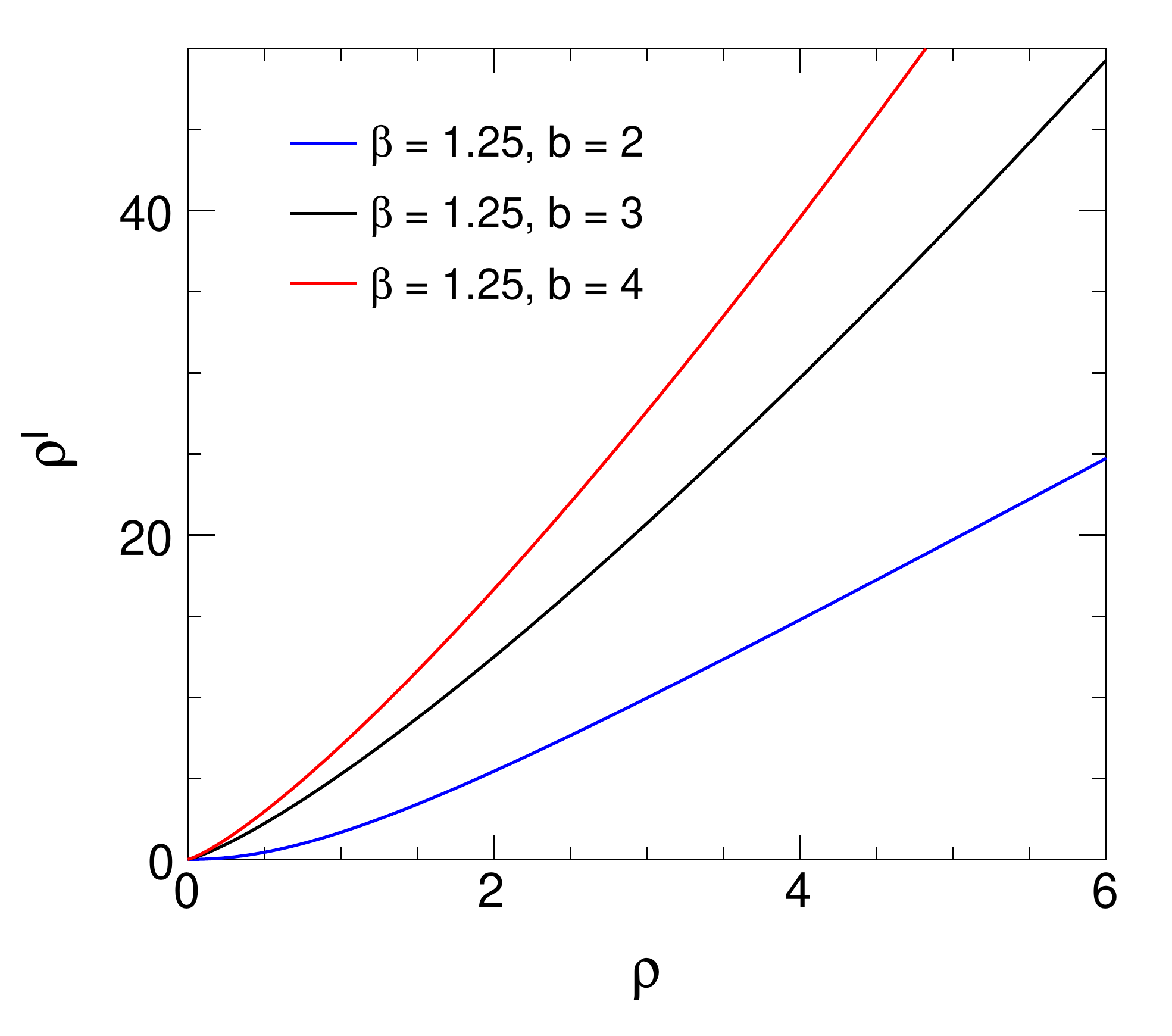}}
\vspace{-0.2cm}
\caption{Relations between interaction and baryonic energy density for 
different values of model parameters $b$ and $\beta$.}
\label{fig1}
\end{figure}	

It is immediately clear from equation \eqref{eq37} that the interaction energy 
density $\rho^I$ is highly model dependent and depends on the positive as well 
as finite values of $b$ and $\beta$. However, it does not depend on the 
geometry of spacetime directly for this model. We know that the galactic 
rotation curves show the linear increase of rotational velocity from the 
center of the galaxy and then remain constant giving flat curves in the halo 
region at large distance from the center, where a very low amount of baryonic 
mass is observed \cite{MPS,J9,Fra}. The interpretation of rotation curves at 
large distances is generally given by postulating the existence of DM in this 
region. Here, we have the relation \eqref{eq37}, generated by our 
$f(\mathcal{R},T)$ gravity model \eqref{eq35} that can explain the rotation 
curves without the need of DM. This would be clear from the graphical analysis 
of the relations \eqref{eq37}. For this we proceed by plotting interaction 
mass density $\rho^I$ against the baryonic mass density $\rho$ using this 
relation as shown in Fig.~\ref{fig1}. The figure indicates two plausible 
requirements: (i) $\rho^I \ge 0$ and (ii) $\rho \ll \rho^I$ for the positive 
values of $b$ and $\beta$.  		

In particular the requirement (ii) tells us about the domination of 
geometrically induced interaction mass density over the baryonic mass density 
at large distances from the centers of galaxies. The geometrical mass 
associated with this interaction mass density may be responsible for the 
constancy of tangential velocity in galactic halo. Furthermore, significant 
domination of interaction density over the baryonic mass density for larger 
values of $b$ and $\beta$ is seen from Fig.~\ref{fig1}. Also, the radial 
interaction pressure in the model \eqref{eq35} takes the form:
\begin{equation}
P_r^I = -\, \frac{b}{16\,\pi\,G}\,(-T)^{\beta}.
\label{eq38}  
\end{equation}
Using expressions \eqref{eq37} and \eqref{eq38} for $\rho^I$ and $ P_r^I$ 
respectively, we obtain,	
\begin{equation}
\rho^I + P_r^I = \frac{b\,\beta}{8\,\pi\,G}\,\rho^{\beta}.
\label{eq39}
\end{equation}
From Fig.~\ref{fig1} we may assume that contribution of the baryonic mass 
density to the total mass density exterior to the galactic disk is very less 
in comparison to the interaction mass density, consequently equation 
\eqref{eq31} can be written using equation \eqref{eq39} as	
\begin{equation}
e^{2\nu} = \frac{\delta\, a}{b\,\beta\,r^2\,\rho^{\beta}}.
\label{eq40}
\end{equation}
Accordingly $g_{00}$ component of the metric coefficient will be
\begin{equation}
 e^{2\mu} =  \frac{b\, \beta\,\rho^{\beta}\,r^{\delta\,+ 2}}{\delta\, a\,h^{\delta}}.
\label{eq41}	
\end{equation}	
Equations \eqref{eq40} and \eqref{eq41} are the explicit forms of the metric 
coefficients in our specified $f(\mathcal{R},T)$ model \eqref{eq35}.
In the following section we will find the rotation curve according to variation of $r$ from the galactic center for the model \eqref{eq35}. 

\section{Galactic Rotation Curves in a Spherically Symmetric Spacetime}
\label{sec5}
\subsection{Orbital motion of a test particle in stable circular orbit}
The tangential velocity of a star in an orbit of radius $r$ around the 
galactic center significantly reflects the matter distribution of the galaxy 
within it. Therefore, we consider a test particle (a star) moving in a static 
spherically symmetric spacetime described by the metric \eqref{eq15}, which 
follows a time-like geodesic confined to the equatorial plane, 
$\theta = \pi/2$. The general geodesic equation that describes the motion of 
such particles has the following form \cite{M14}: 
\begin{equation}
\frac{d^2x^\sigma}{d\tau^2}\, +\, \Gamma_{\mu\nu}^{\sigma} \frac{dx^\mu}{d\tau} \,\frac{dx^\nu}{d\tau} = 0,
\label{eq42}
\end{equation}
where $\tau$ is the affine parameter along the geodesic. In such time-like 
geodesic it represents the proper time. The connection coefficient 
$\Gamma_{\mu\nu}^{\sigma}$ is defined as
\begin{equation}
\Gamma_{\mu\nu}^{\sigma} = \,\frac{1}{2}\,g^{\sigma\psi} \left[\frac{\partial g_{\mu\psi}}{\partial x^\nu} + \frac{\partial g_{\nu\psi}}{\partial x^\mu}-\frac{\partial g_{\mu\nu}}{\partial x^\psi}\right].
\label{eq43}
\end{equation}  
Motion in the equatorial plane ($\theta = \pi/2 \rightarrow \dot{\theta} = 
\ddot{\theta} = 0$) immediately results in three equations of motion for the 
coordinates $t, r, \varphi$ respectively corresponding to the non-vanishing 
components of $\Gamma_{\mu\nu}^{\sigma}$ as follows:
\begin{equation}
\frac{d^2t}{d\tau^2}\, +\,2\,\Gamma_{0 1}^{0}\,\frac{dt}{d\tau} \,\frac{dr}{d\tau} = 0,
\label{eq44}
\end{equation}
\begin{equation}
\frac{d^2r}{d\tau^2}\,+\,\Gamma_{0 0}^{1} \left(\frac{dt}{d\tau}\right)^2 \, +\,\Gamma_{1 1}^{1}\left(\frac{dr}{d\tau}\right)^2\, +\,\Gamma_{3 3}^{1} \left(\frac{d\varphi}{d\tau}\right)^2= 0,
\label{eq45}
\end{equation}
\begin{equation}
\frac{d^2\varphi}{d\tau^2}\,+\,2\,\Gamma_{1 3}^{3}\,\frac{dr}{d\tau}\, \frac{d\varphi}{d\tau} = 0,
\label{eq46}
\end{equation}
where we take $x^0 = t,\, x^1 = r,\, x^2 = \theta$ and $x^3 = \varphi$. For a 
stable circular orbit, after insertion of non-zero values of the connection 
coefficients, equations \eqref{eq44} and \eqref{eq46} take the forms 
respectively as 
$$\frac{d}{d\tau}\left(e^{2\mu}\frac{dt}{d\tau}\right) = 0 \;\;\; \text{and} \;\;\; \frac{d}{d\tau}\left(r^2\,\frac{d\varphi}{d\tau}\right) = 0.$$
The integration of above two equations lead two constants of motion along the 
geodesic, viz.~energy, $E = e^{2\mu}\frac{dt}{d\tau}$ and angular momentum,
$l = r^2\,\frac{d\varphi}{d\tau}$. But at distances far from the spherically 
symmetric body, the spacetime is almost flat and the field is weaker. At this 
weak field limit GR agrees with the Newtonian gravity \cite{rj} and the 
particle moves with a considerably slower speed relative to the speed of 
light. In this low speed approximation, all the spatial components of 
particle's four velocity $u^{k} = \frac{dx^{k}}{d\tau}$, where $k = 1, 2, 3$ 
are dominated by its time component $u^0 = \frac{dx^0}{d\tau}$ \cite{af}. 
Thus, for such a sufficiently slow moving particle, 
$u^{k} = \frac{dx^{k}}{d\tau} \ll \frac{dx^0}{d\tau}$, and the proper time 
$\tau$ may be approximated to the coordinate time $t$ and the four velocity 
$\frac{dx^{\sigma}}{d\tau} (\frac{dx^0}{dt}, \frac{dx^1}{dt}, \frac{dx^2}{dt}, 
\frac{dx^3}{dt})$ as $(1, 0, 0, 0)$ \cite{wa}. With this approximation none of 
the spatial component of the four velocity will appear in the geodesic 
equation \cite{wa} and hence equation \eqref{eq42} takes the form as
\begin{equation}
\frac{d^2x^\sigma}{dt^2}\,+\,\Gamma_{0 0}^{\sigma}  = 0.
\label{eq47}
\end{equation}
Here the connection $\Gamma_{0 0}^{\sigma}$ can be calculated from the 
metric \eqref{eq15} using the relation \eqref{eq43} as
$$\Gamma_{0 0}^{\sigma} = \,\frac{1}{2}\,g^{\sigma\psi}\left[\frac{\partial g_{0 \psi}}{\partial x^0} + \frac{\partial g_{0 \psi}}{\partial x^0} - \frac{\partial g_{0 0}}{\partial x^\psi}\right] = -\,\frac{1}{2} g^{\sigma\psi} \frac{\partial g_{0 0}}{\partial x^\psi}.$$ 
Now, the equation of motion for the radial component will be 
\begin{equation}
	\frac{d^2r}{dt^2}\, +\,\Gamma_{0 0}^{1}  = 0.
	\label{eq48}
\end{equation}
In the Newtonian limit, the radial or centripetal acceleration $a$ of the 
test particle moving with the tangential (circular) velocity $v$ in an orbit 
of radius $r$ is given as 
\begin{equation}
a = -\, \frac{v^2}{r}.
\label{eq49}
\end{equation}
Equations \eqref{eq48} and \eqref{eq49} together give
$ v^2 = r\, \Gamma_{0 0}^{1}$,
where after substitution of corresponding metric elements of the metric 
\eqref{eq15} we obtain $\Gamma_{0 0}^{1} = \mu'e^{2\mu-2\nu}$ and hence
\begin{equation}
v^2 = r \mu'e^{2\mu-2\nu}.
\label{eq50}
\end{equation}
This equation relates the tangential velocity of a test particle to the radial 
coordinate $r$, the $g_{00}$ and $g_{11}$ components of the metric 
\eqref{eq15}.

The metric coefficients $e^{2\nu}$ and $e^{2\mu}$ in terms of $r$ using 
equation \eqref{eq33} can be rewritten from equations \eqref{eq40} and 
\eqref{eq41} in the following forms:
\begin{equation}
e^{2\nu} =\,\frac{\delta\, a}{b\,\beta\,r^{-\,\lambda\beta+2}}\;\;\;\text{and}
\;\;\; e^{2\mu} =  \frac{b\, \beta\,r^{-\,\lambda\beta+\delta\,+2}}{\delta\, a\,h^{\delta}}.
\label{eq51}
\end{equation}
Next, we have to find $\mu'$ to express the tangential velocity of the 
particle in terms of the model parameters. It may be obtained from the 
differentiation of the second relation of \eqref{eq51} with respect to $r$ as 
\begin{equation}
\mu' = \frac{1}{2\,e^{2\mu}}\left[\,\frac{b\,\beta\,(2 + \delta - \lambda\, \beta)\,r^{\,1-\lambda\,\beta + \delta}}{h^{\delta}\, a\, \delta} \right]. 
\label{eq52}
\end{equation}
Thus, insertion of relations \eqref{eq51} and equation \eqref{eq52} in 
equation \eqref{eq50} we get the tangential velocity of a test particle 
rotating around a galaxy in terms of radial coordinate $r$ for our assumed 
$f(\mathcal{R},T)$ gravity model \eqref{eq35} as   
\begin{equation}
v^2 = \frac{b^2\beta^2 (2+\delta-\lambda \beta)r^{4+\delta-2 \lambda \beta}}{2\,h^{\delta}a^{2}\,\delta^2}  
\label{eq53}
\end{equation}
Relation \eqref{eq53} is the main equation for fulfilment of the aim of the 
study. In the following we will compare this result with some observed rotation 
velocity data to test the viability of the model considered for the study. 

\begin{table}[h!]
\centering
\caption{Sample of observed Galaxies.} 
\vspace{0.2cm}
\scalebox{0.95}{
\begin{tabular}{c c c c c c c c c c c c}  \hline\\[-8pt]
& Galaxy Name& ~~~Type & ~~~Distance ($D$) &  ~~~Luminosity  &  ~~~Scale length ($h$) & Data Sources & \\
& & &   ~~~(Mpc)      &  ~~~$L_{B}$     &    ~~~(kpc)~~~ & $v$~~~ $D$~~~ $L_B$~~~ $h$ &  \\[2pt] \hline \hline\\[-8pt]
& F 583-1  & LSB  & 35.4 & 0.064 & 1.6 & \cite{spark} \cite{spark} \cite{pd} \cite{pd} & \\ [7pt]
& DDO 154 &  LBS & 4.04 &  0.008& 0.54& \cite{spark} \cite{spark} \cite{pd} \cite{pd} & \\ [7pt]
 & UGC 128 & LSB  & 64.5 & 0.597 & 6.9 & \cite{spark} \cite{spark} \cite{pd} \cite{pd} & \\ [7pt]
& UGC 1230& LSB  & 53.7 & 0.366 & 4.7 & \cite{spark} \cite{spark} \cite{pd} \cite{pd} & \\ [7pt]
& UGC 1281& LSB  & 5.1  & 0.017 & 1.6 & \cite{spark} \cite{np}    \cite{np} \cite{pd} & \\ [7pt]
& UGC 6446& LSB  & 12.0 & 0.263 & 1.9 & \cite{spark} \cite{spark} \cite{pd} \cite{pd} & \\ [7pt]
& NGC 0247& LSB  & 3.7  & 0.512 & 4.2 & \cite{spark} \cite{spark} \cite{pd} \cite{pd} & \\ [7pt]
& NGC 0300& LSB  & 2.08 & 0.271 & 2.1 & \cite{spark} \cite{spark} \cite{pd} \cite{pd} & \\ [7pt]
& NGC 1003& LSB  & 11.4 & 1.480 & 1.9 & \cite{spark} \cite{spark} \cite{pd} \cite{pd} & \\ [7pt]
& NGC 2976& LSB  & 3.58 & 0.201 & 1.2 & \cite{spark} \cite{spark} \cite{pd} \cite{pd} & \\ [7pt]
& NGC 2403& HSB  & 3.16 & 1.647 & 2.7 & \cite{spark} \cite{spark} \cite{pd} \cite{pd} & \\ [7pt]
& NGC 2998 & HSB  & 68.1 & 5.186 & 4.8 & \cite{spark} \cite{spark} \cite{pd} \cite{pd} & \\ [7pt]
& NGC 3198& HSB  & 13.8 & 3.241 & 4.0 & \cite{spark} \cite{spark} \cite{pd} \cite{pd} & \\[7pt]
& NGC 3521 & HSB  & 7.7  & 2.048 & 3.3 & \cite{spark} \cite{spark} \cite{pd} \cite{pd} & \\ [7pt]
& UGC 3580& HSB  & 20.7 &10.12  & 2.4 & \cite{spark} \cite{spark} \cite{PL} \cite{PL} & \\ [7pt]
& NGC 4088& HSB  & 18.0 & 2.957 & 2.8 & \cite{spark} \cite{spark} \cite{pd} \cite{pd} & \\ [7pt]
& NGC 4183& HSB  & 18.0 & 1.042 & 2.9 & \cite{spark} \cite{spark} \cite{pd} \cite{pd} & \\ [7pt]
& NGC 5585 & HSB  & 7.1  & 0.333 & 2.0 & \cite{spark} \cite{spark} \cite{pd} \cite{pd} & \\ [7pt]
& NGC 7793 & HSB  & 3.61 & 0.910 & 1.7 & \cite{spark} \cite{spark} \cite{pd} \cite{pd} & \\[2pt] \hline \hline
\end{tabular}}\\[5pt]
LSB $\rightarrow$ Low surface brightness; HSB $\rightarrow$ High surface 
brightness.
\label{table1}
\end{table}

\subsection{Fitting of Rotation Curve}
\begin{figure}[!h]
\centerline{
\includegraphics[scale = 0.275]{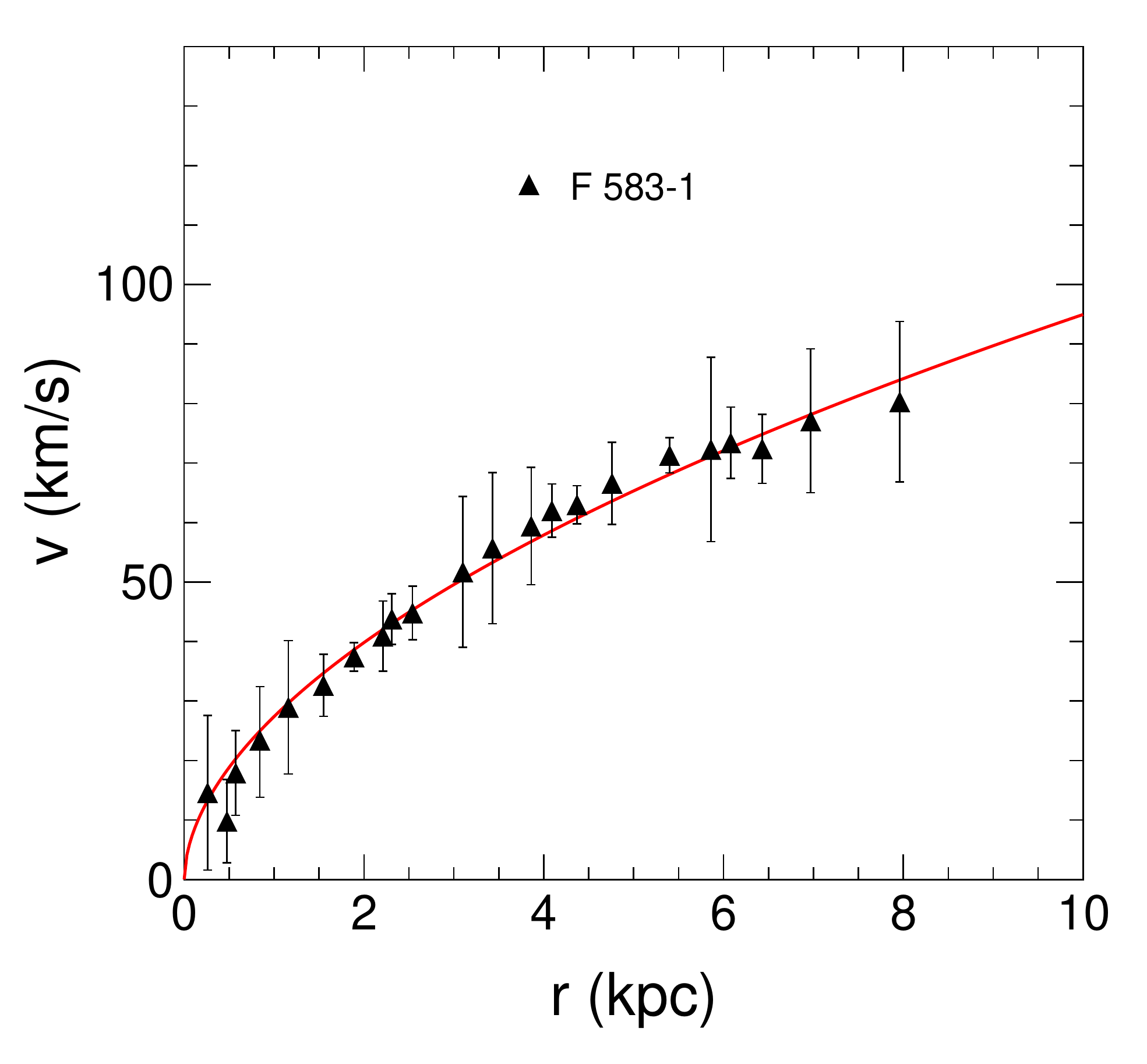}\hspace{.09cm}
\includegraphics[scale = 0.275]{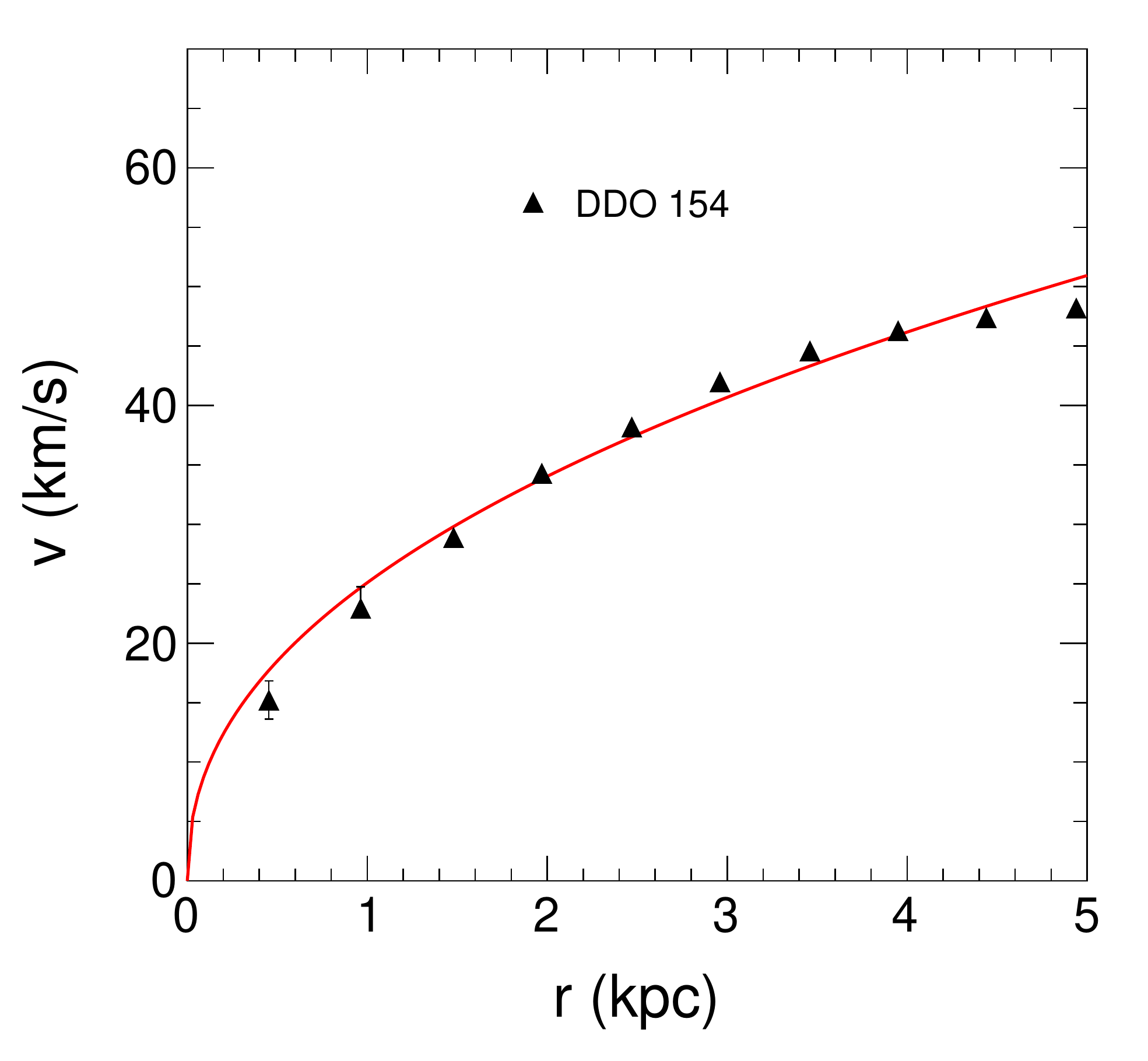}\hspace{.09cm}
\includegraphics[scale = 0.275]{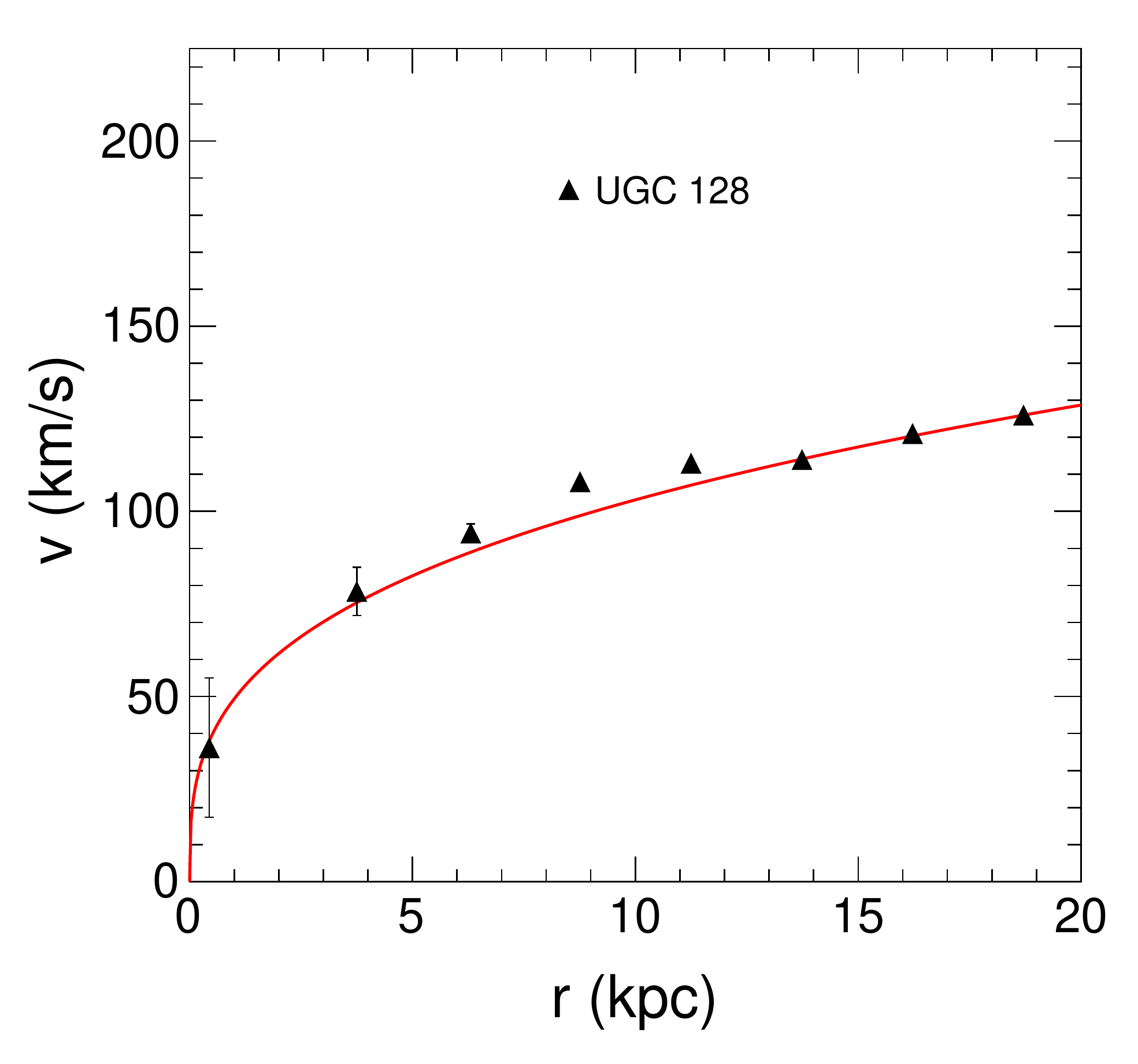}}\hspace{.09cm}
\includegraphics[scale = 0.275]{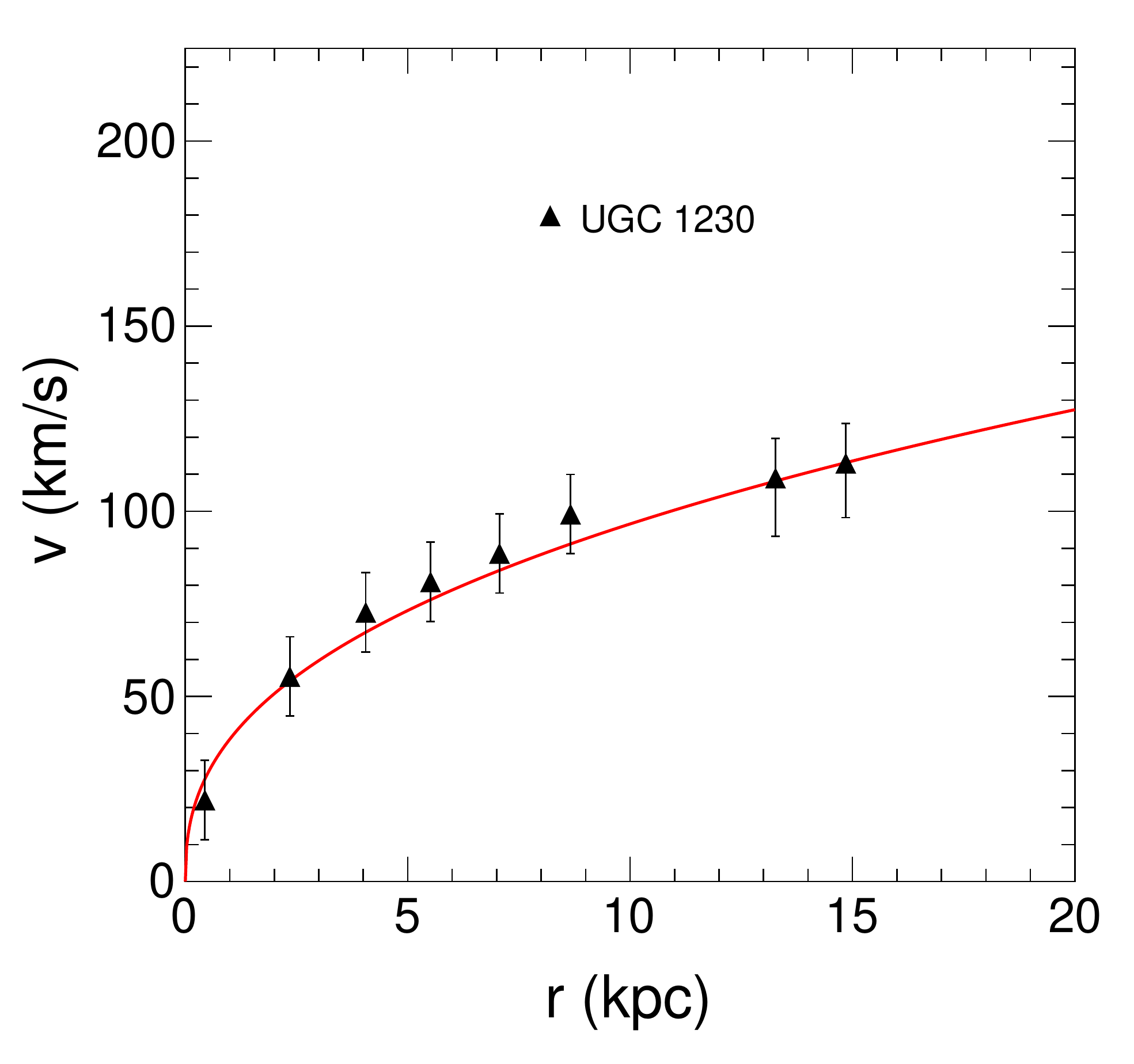}\hspace{.09cm}
\includegraphics[scale = 0.275]{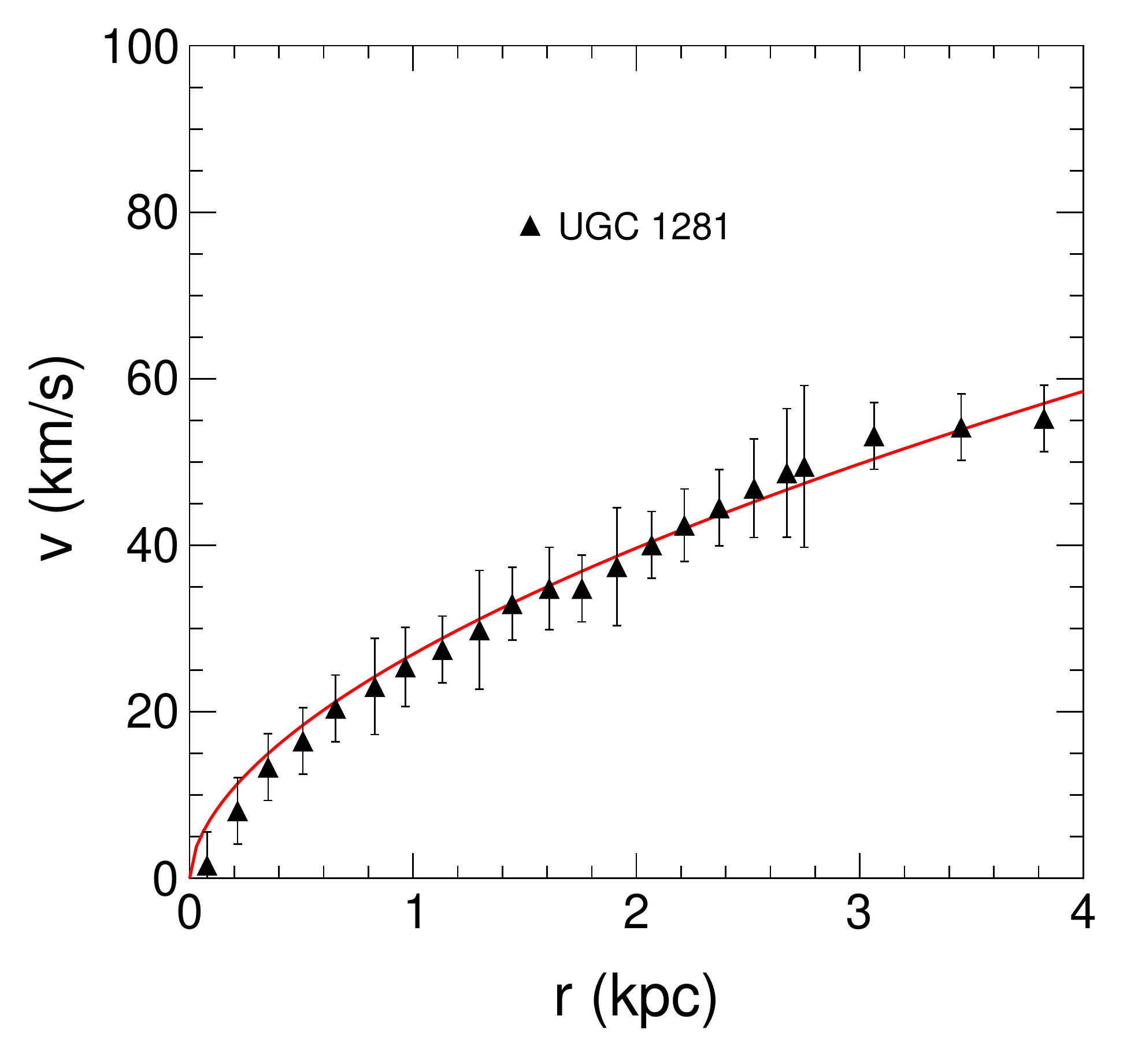}\hspace{.09cm}
\includegraphics[scale = 0.275]{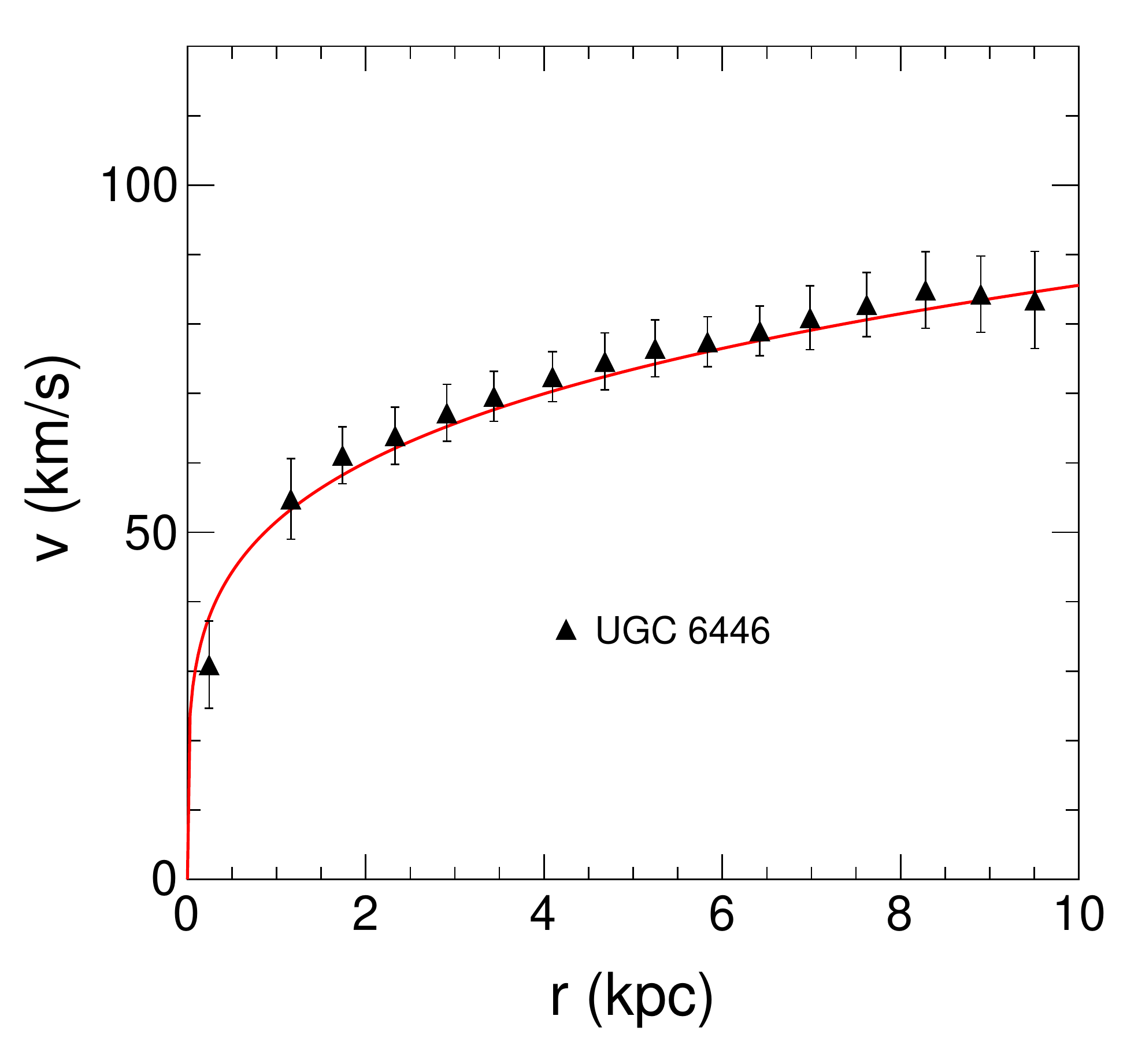}\hspace{.09cm}
\includegraphics[scale = 0.275]{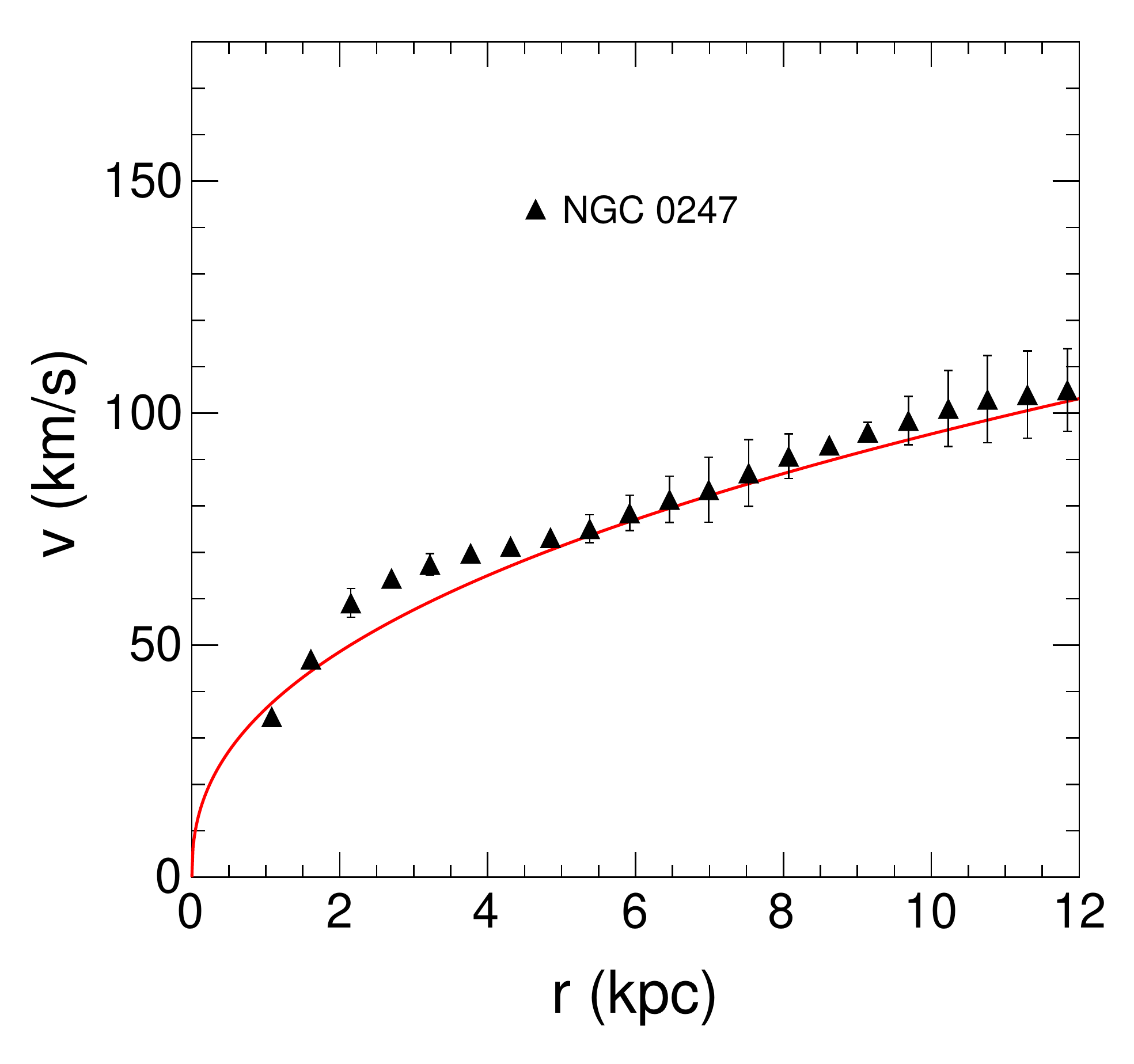}\hspace{.09cm}
\includegraphics[scale = 0.275]{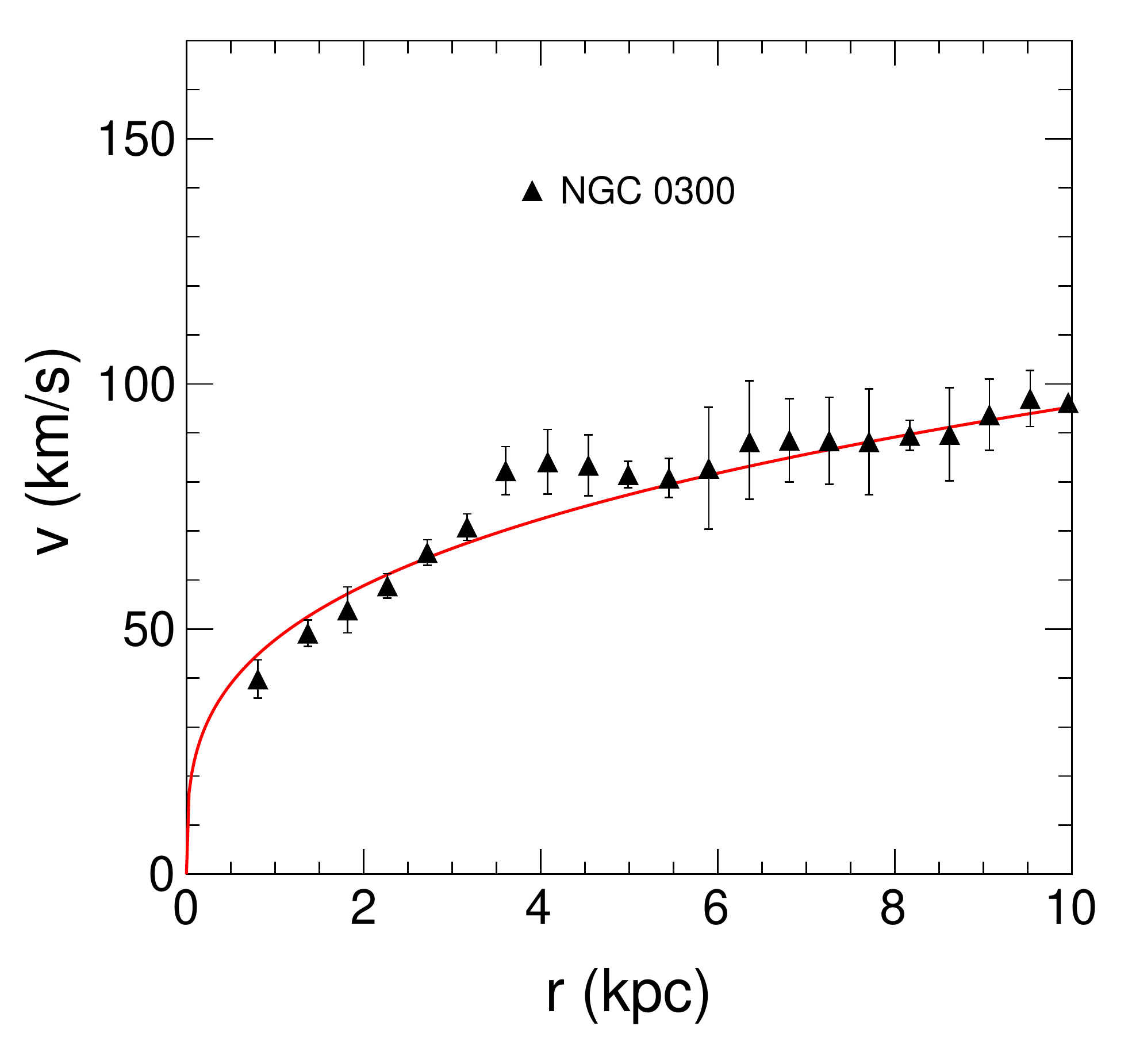}\hspace{.09cm}
\includegraphics[scale = 0.275]{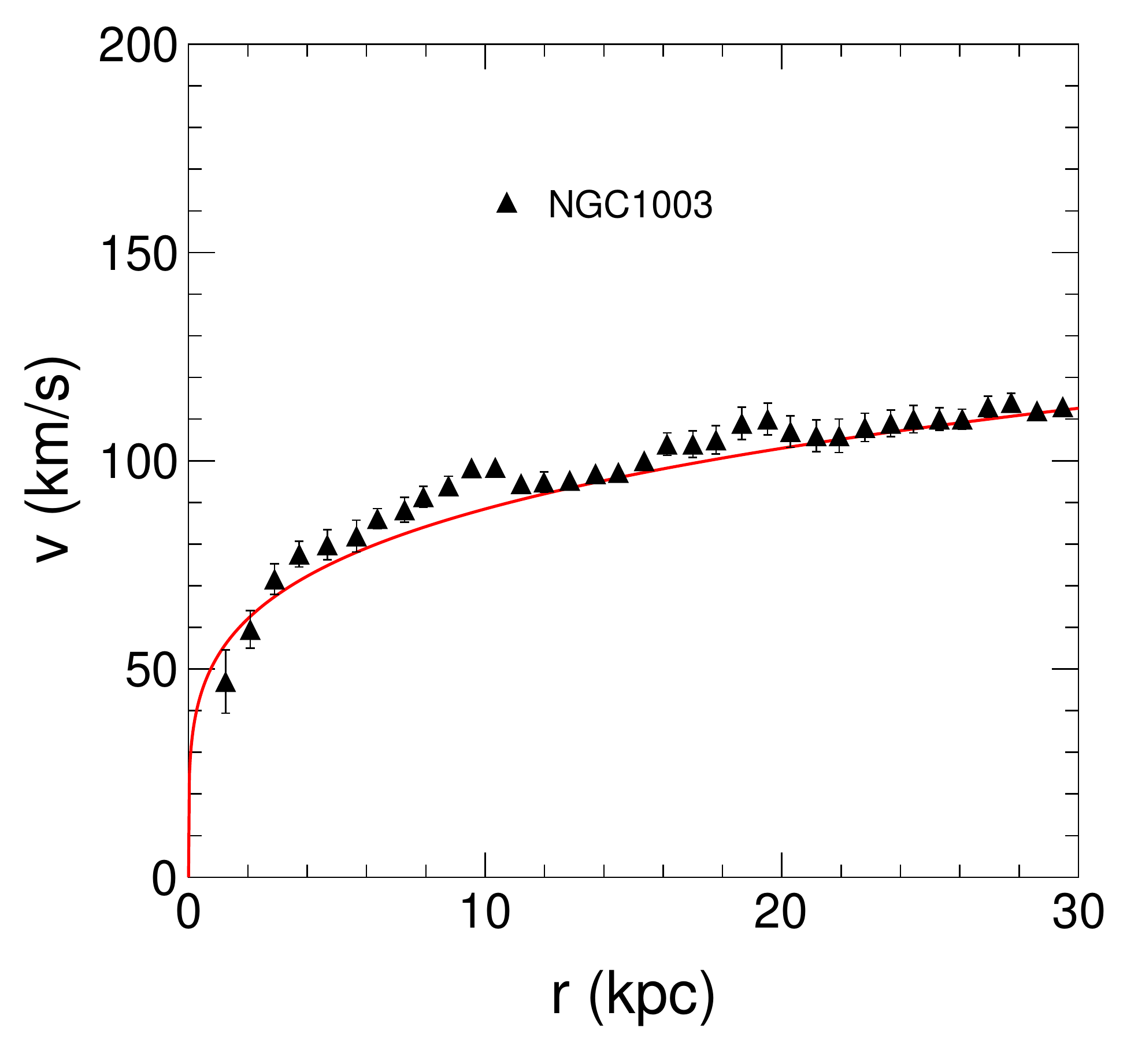}\hspace{.09cm}
\includegraphics[scale = 0.275]{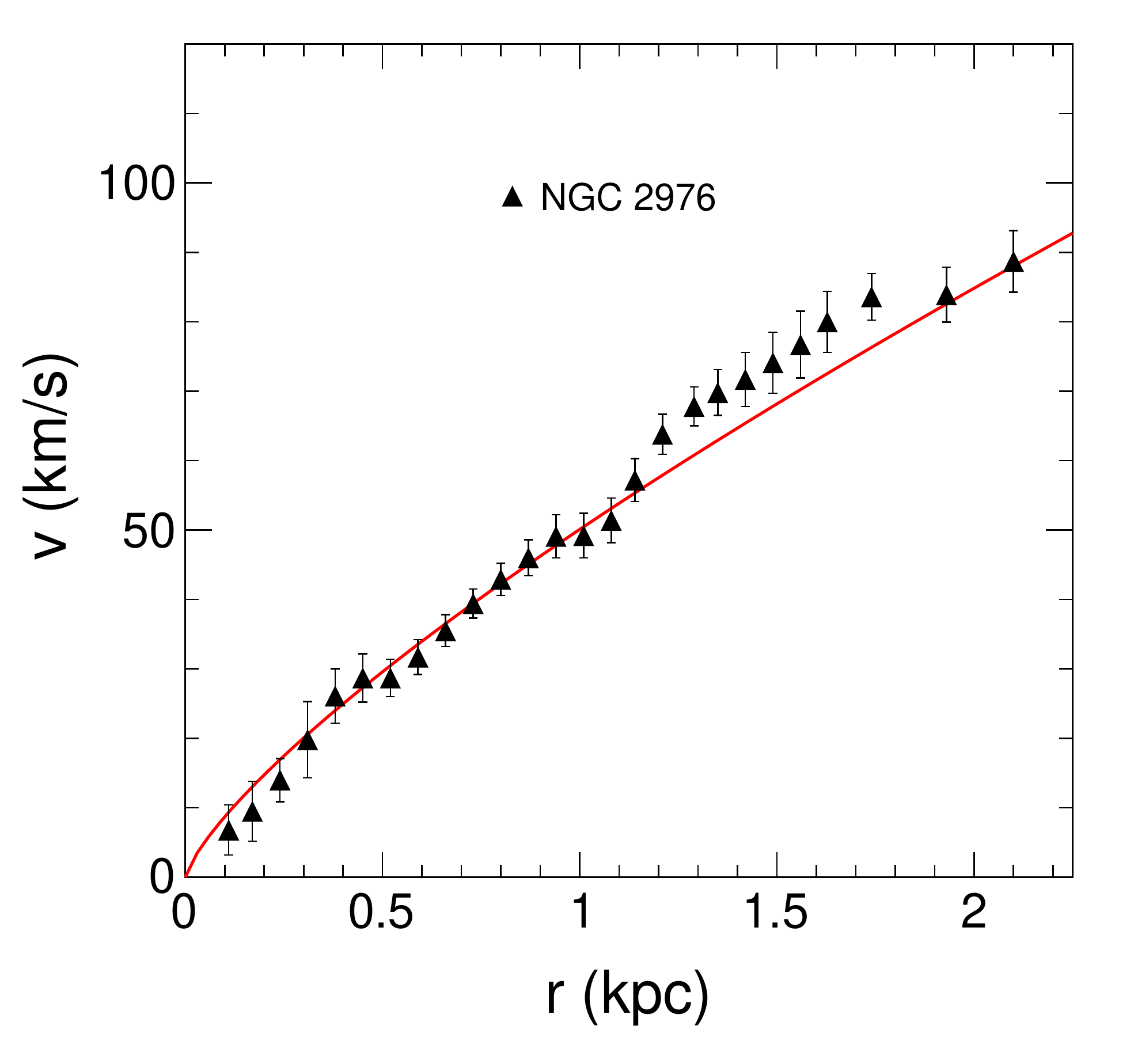}\hspace{.09cm}
\vspace{-0.2cm}	
\caption{Fitting of rotation curves generated from the model \eqref{eq35} to 
rotation velocities of samples of 10 LSB galaxies with their quoted errors. 
The data points are observational values of rotational velocities extracted 
from Ref.~\cite{spark}.}
\label{fig2}
\end{figure}

\begin{figure}[!h]
\centerline{
\includegraphics[scale = 0.275]{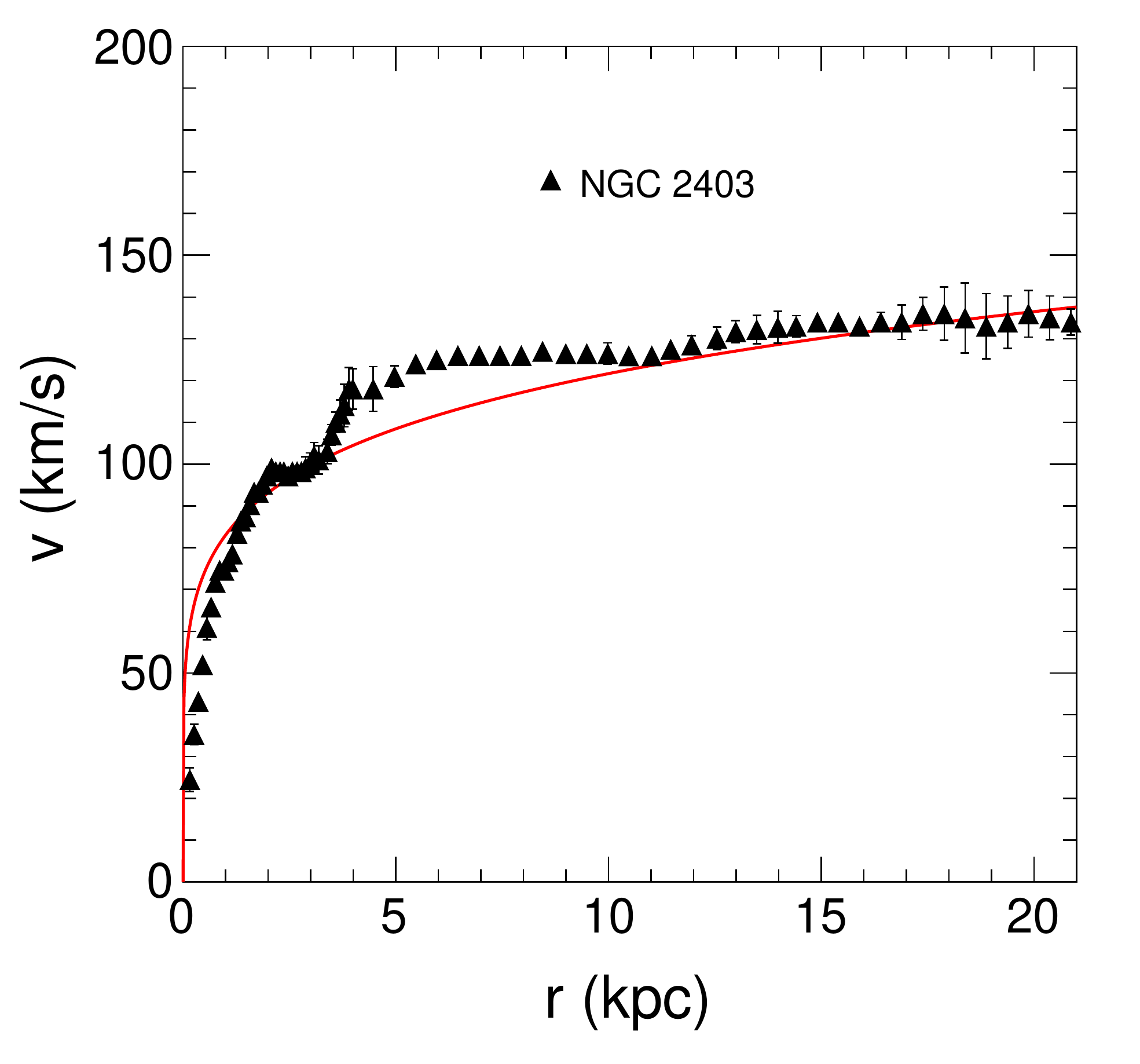}\hspace{.09cm}
\includegraphics[scale = 0.275]{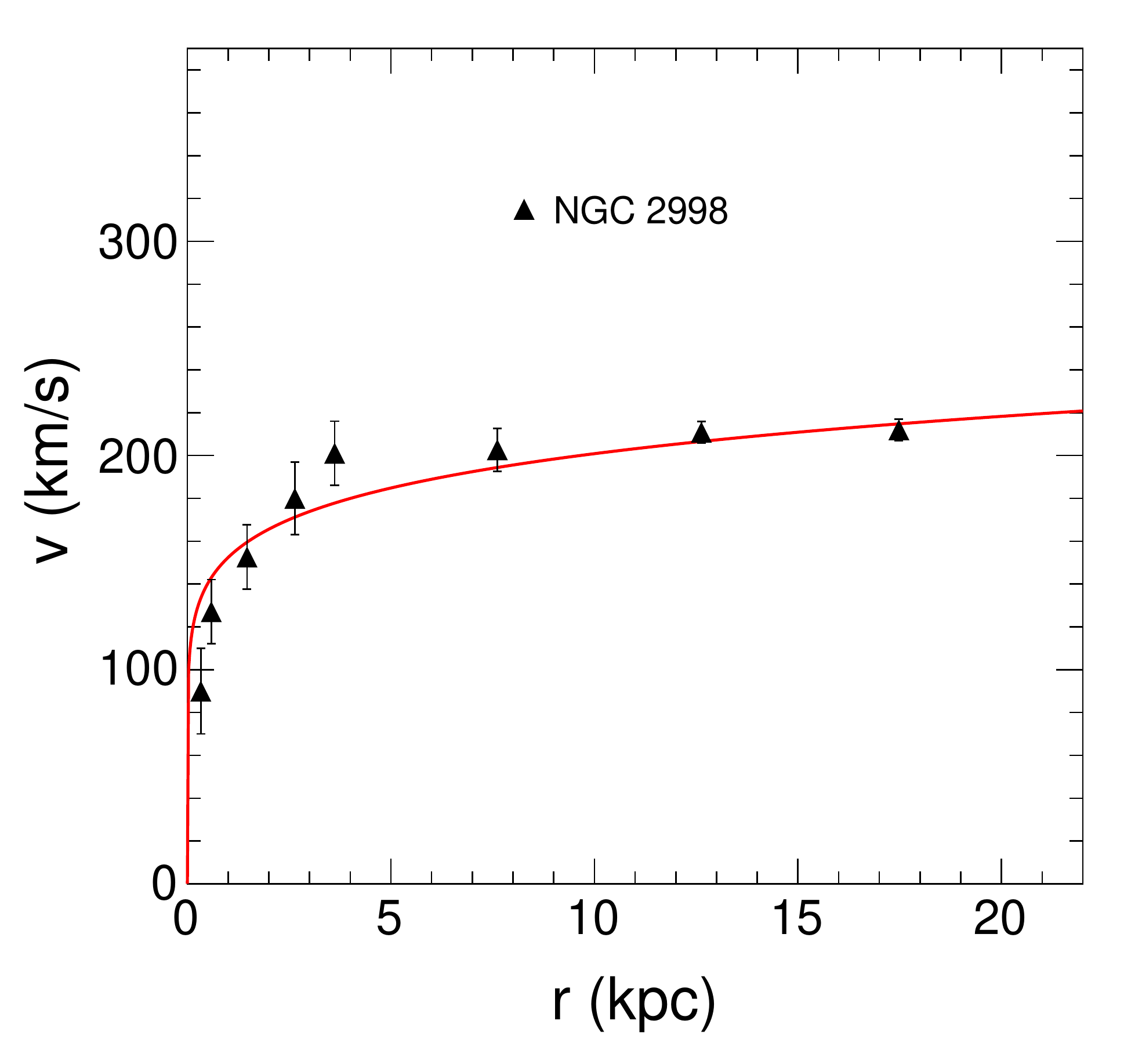}\hspace{.09cm}
\includegraphics[scale = 0.275]{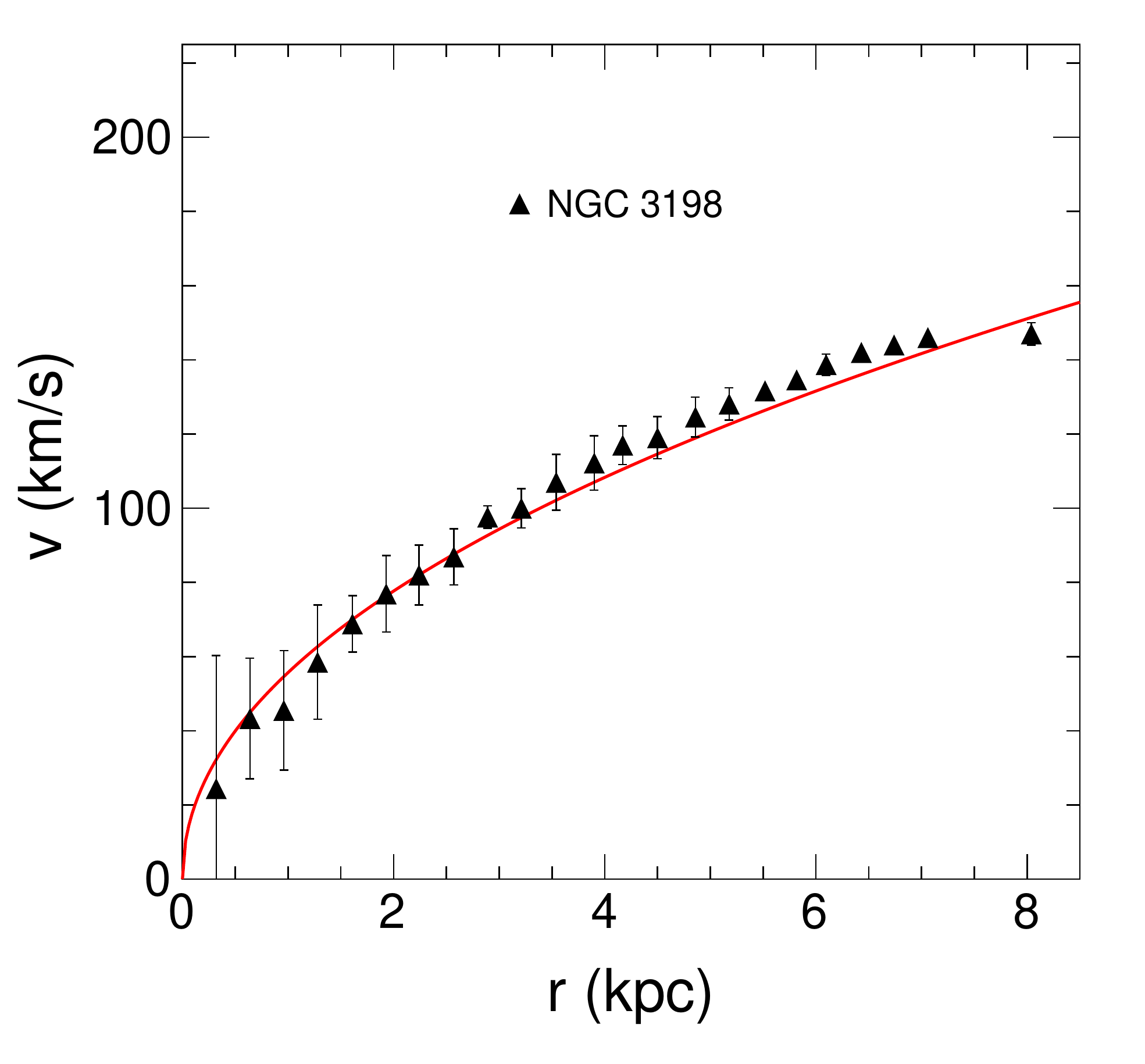}\hspace{.09cm}}
\includegraphics[scale = 0.275]{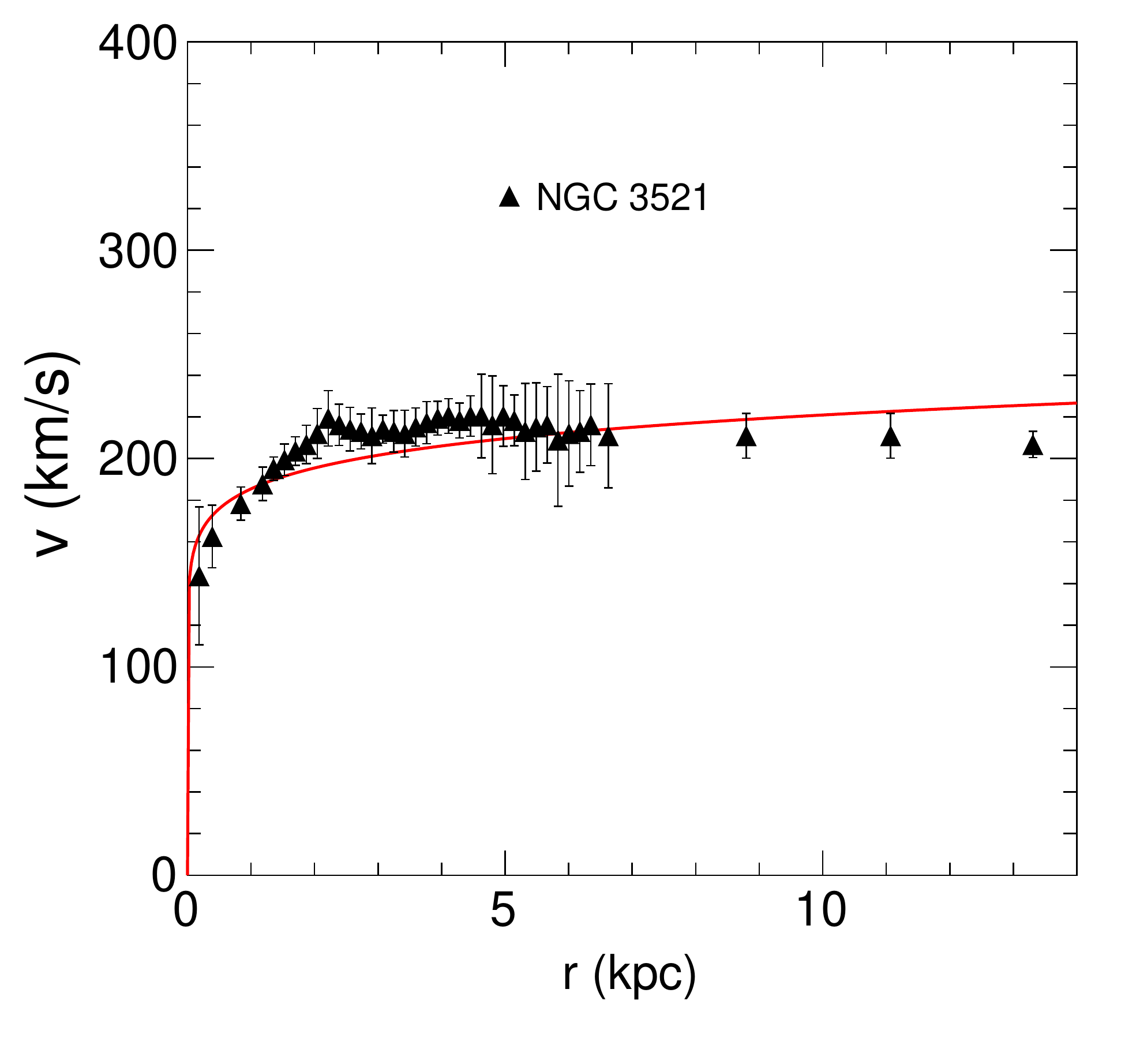}\hspace{.09cm}
\includegraphics[scale = 0.275]{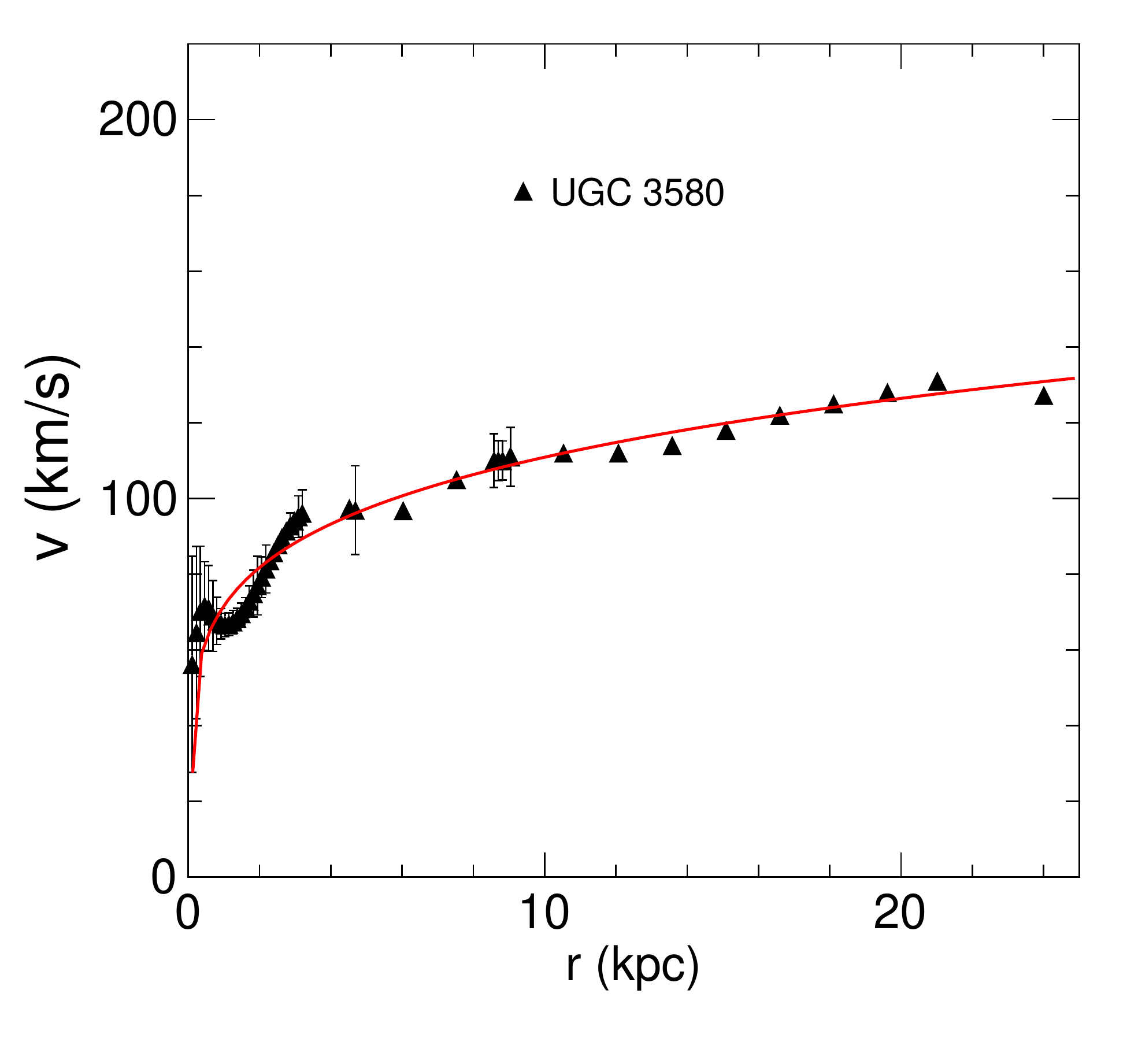}\hspace{.09cm}
\includegraphics[scale = 0.275]{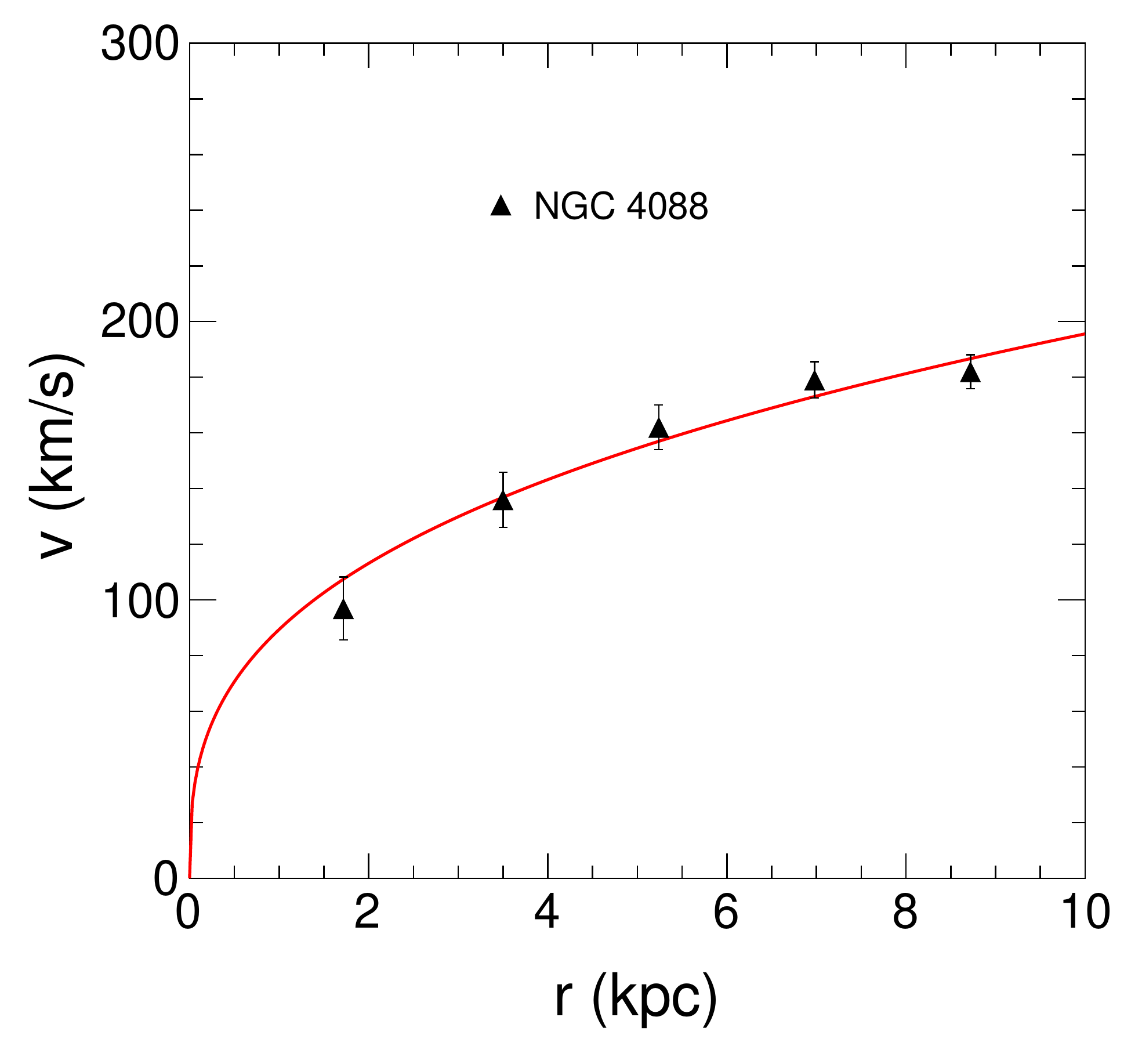}\hspace{.09cm}
\includegraphics[scale = 0.275]{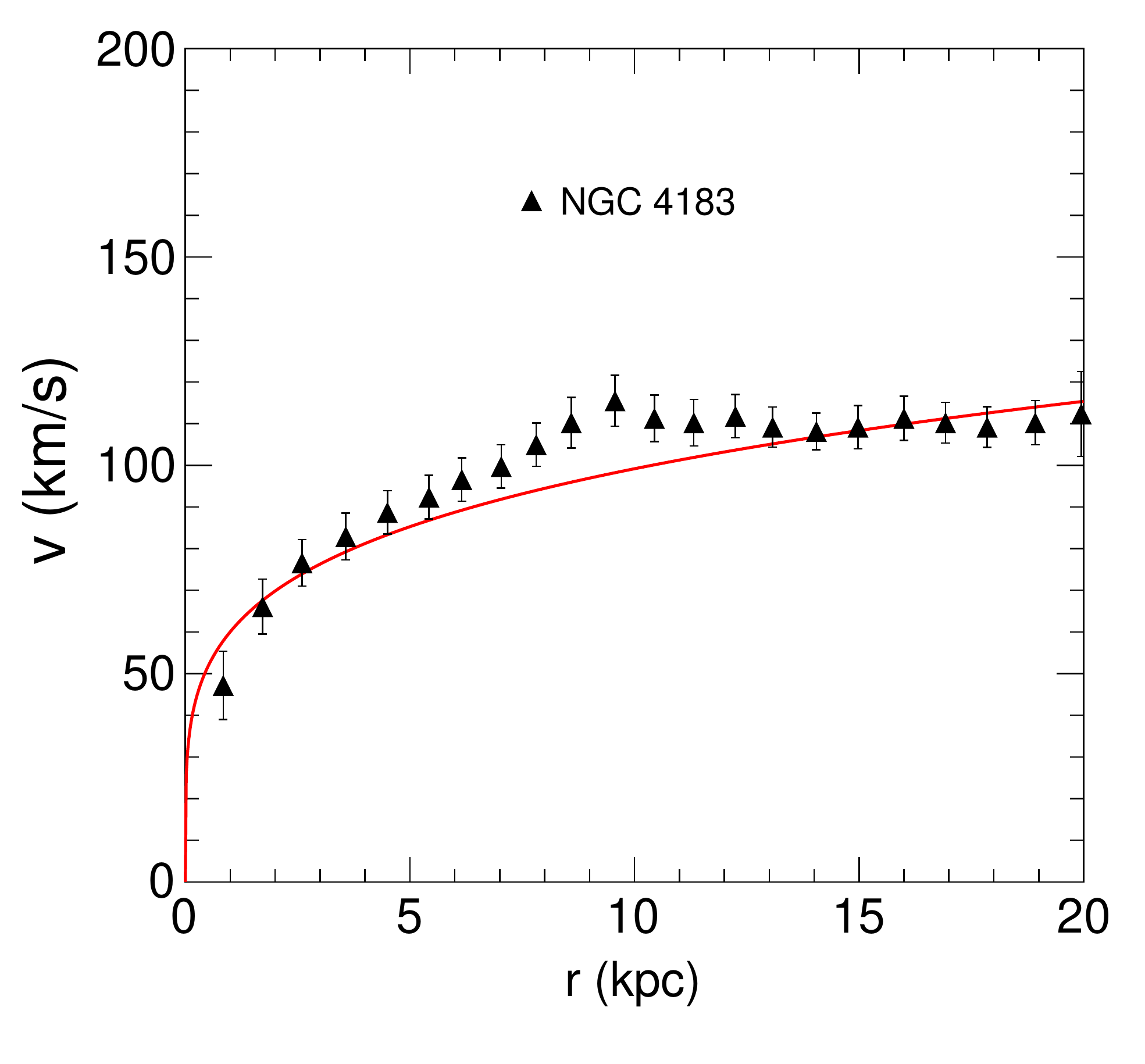}\hspace{.09cm}
\includegraphics[scale = 0.275]{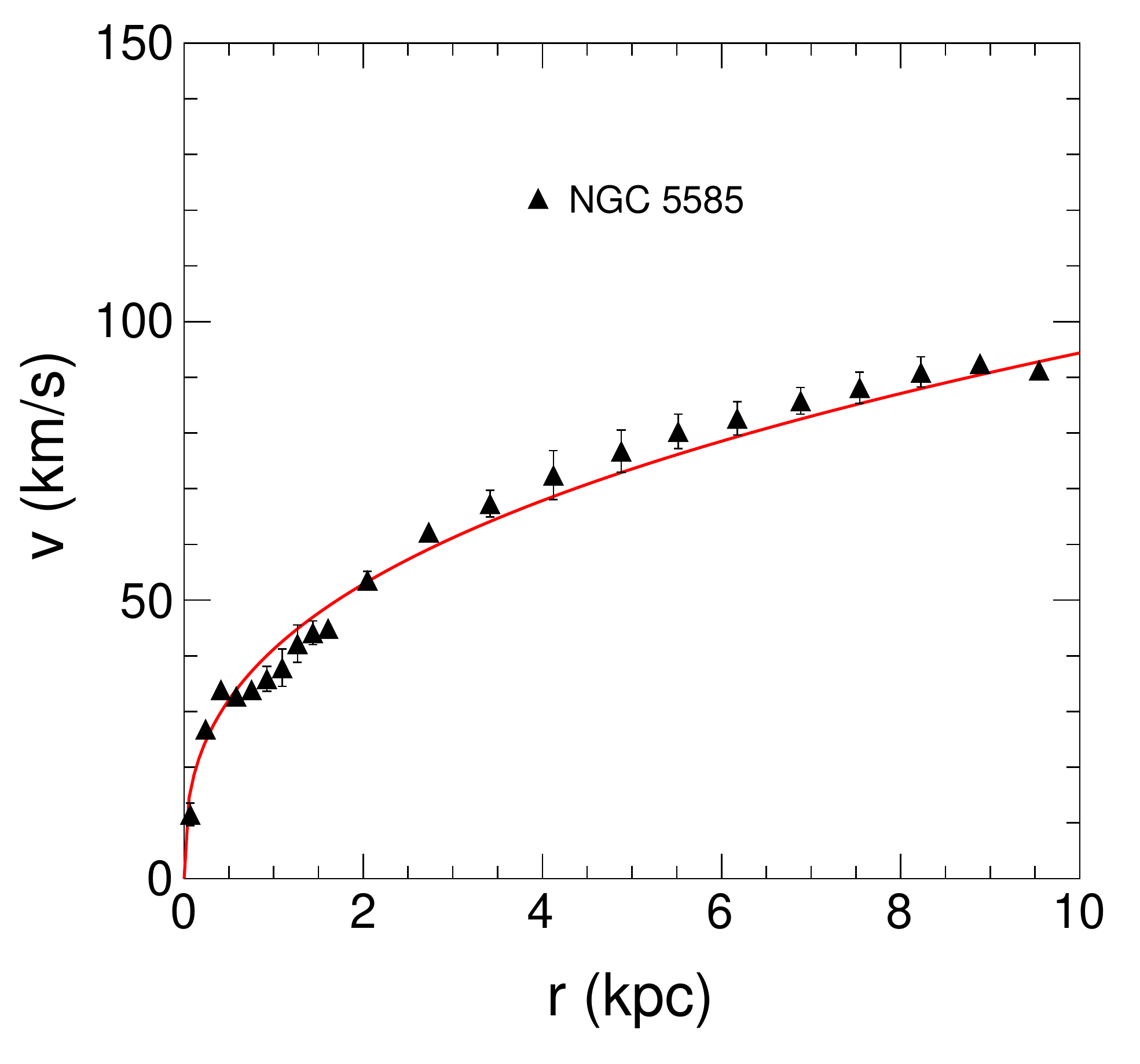}\hspace{.09cm}
\includegraphics[scale = 0.275]{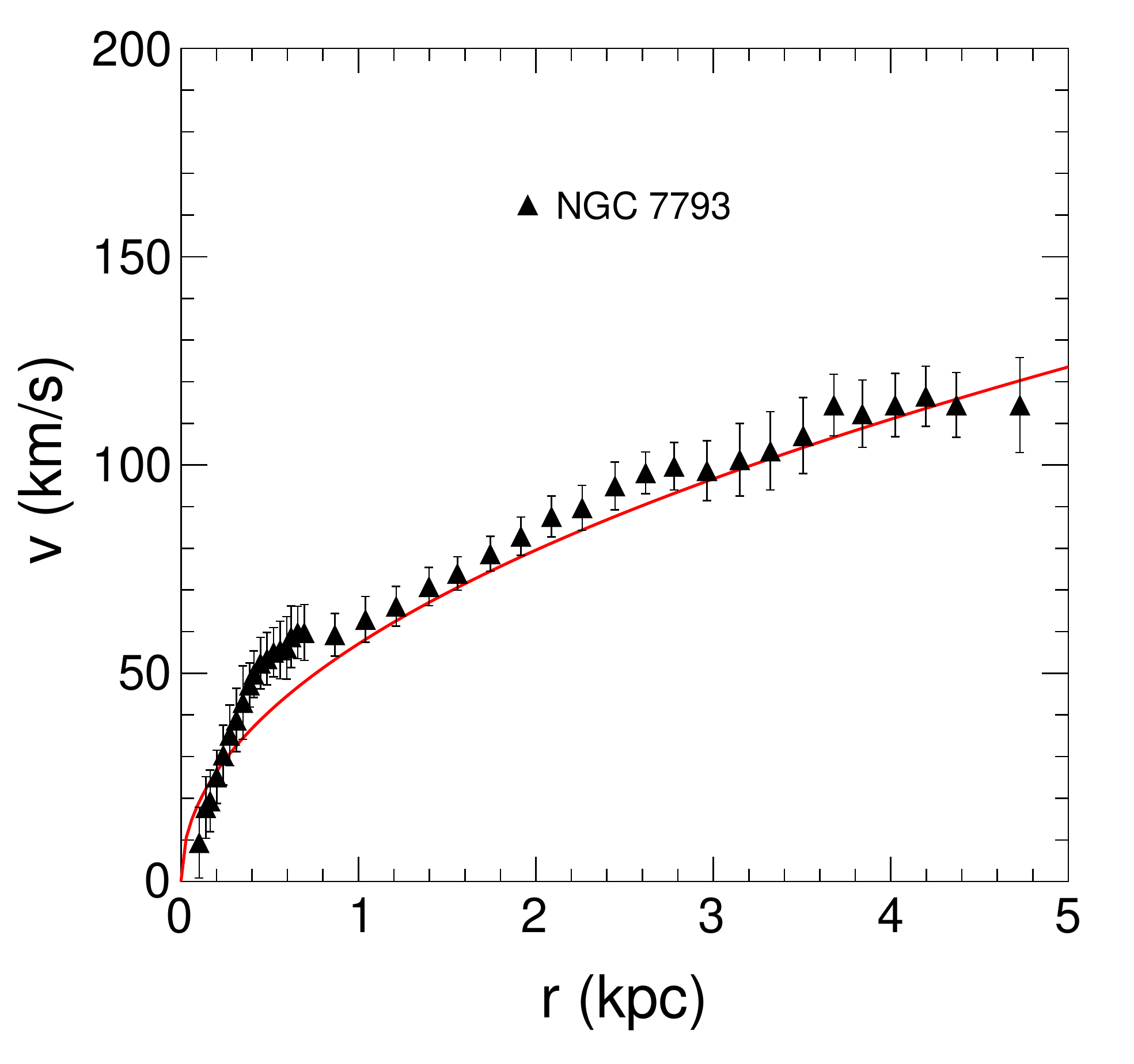}
\vspace{-0.2cm}
\caption{Fitting of rotation curves generated from the model \eqref{eq35} to 
rotation velocities of samples of 9 HSB galaxies with their quoted errors. The 
data points are observational values of rotational velocities extracted from 
Ref.~\cite{spark}.}
\label{fig3}
\end{figure}

We have generated the rotation curves (see Figs.~\ref{fig2} and \ref{fig3}) for 
different values of the model parameters according to equation \eqref{eq53} 
for the test particle moving around the galactic center and test the model by 
fitting the predicted results to observational data of a sample of 19 
galaxies \cite{spark}, 10 LSB and 9 HSB, listed in Table~\ref{table1} with 
the data extracted from 
the references indicated in the table. As different types of galaxies exhibit 
separate morphologies, we chose these few galaxies as typical galaxies to 
represent the morphologies of similar kinds of other galaxies. In 
Table~\ref{table1}, data of the adopted distance $D$ of sample galaxies 
measured in the unit of Mpc, the B-band luminosity $L_B$ and scale length $h$ 
in kpc are also listed. It is seen that for all samples the predicted rotation 
curves are almost in good agreement with observed data showing well fitted 
results between the model and observations. The reduced $\chi^2$ 
($\chi_{red}^2$) value of fitting for each selected galaxy is calculated and 
indicated in the table Table~\ref{table2}. Eventually $\chi_{red}^2$ values 
for galaxies DDO $154$, NGC $2403$ and NGC $5585$ are found to be large 
($\sim 11.1, 8.5$ and $6.6$ respectively) and for galaxies UGC $128$, NGC 
$0247$, NGC $1003$, UGC $3580$, NGC $2998$, UGC $1230$ are high around $3.6, 
3.4, 3.03, 2.5, 2.2, 2.1$ respectively. But the shape of rotation curves are 
not affected by these large or high $\chi_{red}^2$ values, which are depicted 
in Figs.~\ref{fig2} and \ref{fig3}. 
Table~\ref{table2} also shows the best fitted values of model parameters 
for each sample galaxy and also the value of $\delta$, which must take a very 
small value as mentioned after the equation \eqref{eq27}. For all well fitted 
curves the values of $\delta$ are obtained in the order of 
$10^{-6}$ -- $10^{-5}$. Values of parameters $b$ and $\beta$ are not too 
large, $a$ lies in $700$ -- $800$ and the product $\lambda \beta$ 
is in between $1.28$ and $2$. Both the HSB galaxies which have steeply rising 
rotation curves and the LSB galaxies which are supposed to be DM dominated at 
small radii are seen to be well fitted in the model considered for the study.

\begin{table}[!h]
\centering
\caption{Best fit values of $\delta$ and different model parameters.}
\vspace{0.2cm}
\scalebox{0.95}{
\begin{tabular}{c c c c c c c c} \hline\\[-8pt]
Galaxy Name & ~~~~~~$a$~~~~~~ & ~~~~~~$b$~~~~~~ & ~~~~~~$\lambda$~~~~~~ & ~~~~~~$~~~~~~\delta$~~~~~~ & ~~~~~~$\beta$~~~~~~ & ~~~~~~$~~~~~~\lambda\beta$~~~~~~ &~~~~~~$\chi^2_{red}$ ~~~~~\\[2pt] \hline \hline\\[-8pt]
F583-1   & 710  & 0.46  & 2.0 & $9 \times10^{- 6}$  & 0.730 & 1.46 & 0.37 \\ [7pt]
DDO 154  & 750  & 0.40  &1.95 & $ 8 \times10^{- 6}$ & 0.800 & 1.56 & 11.1 \\ [7pt]
UGC 128  & 700  & 0.82  & 2.0 & $8 \times10^{- 6}$  & 0.840 & 1.68 & 3.6 \\ [7pt]
UGC 1230 & 800  & 1.50  & 2.0 & $7 \times10^{- 6}$  & 0.920 & 1.84 & 2.1 \\ [7pt]
UGC 1281 & 700  & 0.20  & 2.0 & $4 \times10^{- 6}$  & 0.640 & 1.28 & 0.28\\ [7pt]
UGC 6446 & 730  & 0.89  & 2.0 & $7 \times10^{- 6}$  & 0.890 & 1.78 & 0.58\\ [7pt]
NGC 0247 & 800  & 0.40  & 2.0 & $5 \times10^{- 6}$  & 0.790 & 1.58 & 3.4 \\ [7pt]
NGC 0300 & 790  & 0.80  & 2.0 & $7 \times10^{- 6}$  & 0.850 & 1.70 & 1.4 \\ [7pt]
NGC 1003 & 710  & 1.15  & 2.0 & $9 \times10^{- 6}$  & 0.890 & 1.78 & 3.03\\ [7pt]
NGC 2403 & 798  & 2.30  & 2.0 & $9.2 \times10^{- 6}$& 0.917 & 1.83 & 8.5 \\ [7pt]
NGC 2998 & 720  & 3.80  & 2.0 & $8 \times10^{- 6}$  & 0.940 & 1.88 & 2.2 \\ [7pt]
NGC 2976 & 700  & 1.58  & 2.0 & $9 \times10^{- 6}$  & 0.870 & 1.74 & 0.94\\ [7pt]
NGC 3198 & 700  & 0.82  & 2.0 & $8 \times10^{- 6}$  & 0.760 & 1.52 & 1.13\\ [7pt]
NGC 3521 & 710  & 5.60  & 2.0 & $8 \times10^{- 6}$  & 0.962 & 1.92 & 0.95\\ [7pt]
UGC 3580 & 700  & 1.30  & 2.0 & $8 \times10^{- 6}$  & 0.895 & 1.79 & 2.5 \\ [7pt]
NGC 4088 & 720  & 1.50  & 2.0 & $9 \times10^{- 6}$  & 0.830 & 1.66 & 0.92\\ [7pt]
NGC 4183 & 710  & 1.30  & 2.0 & $9 \times10^{- 6}$  & 0.891 & 1.78 & 1.8 \\ [7pt]
NGC 5585 & 720  & 0.68  & 2.0 & $8 \times10^{- 6}$  & 0.820 & 1.64 & 6.6 \\ [7pt]
NGC 7793 & 720  & 0.88  & 2.0 & $8 \times10^{- 6}$  & 0.760 & 1.52 & 0.56 \\[2pt] \hline \hline	
\end{tabular}}
\label{table2}	
\end{table}

\section{Conclusions}
\label{sec6}
MTGs provide alternative gravitational approaches in contrast to the GR to 
understand and determine the flatness of rotation curves corresponding to 
stars that are moving around the galactic nucleus at far distances away from 
it. Thus MTGs can be used to give the geometric interpretation of the 
so-called DM. In this present work, we have developed such an approach to 
address the DM problem by considering the $f(\mathcal{R},T)$ modified theory 
of gravity especially in the galactic scale and emphasis is given to the 
possibility for elucidating the rotation curves of spiral galaxies without 
considering the need for the exotic DM. Here, we have employed a minimally 
coupled $f(\mathcal{R},T)$ gravity model of the form $af(R) + bf(T)$, which 
could explain the additional matter content needed for the observational 
agreement with the rotation curves of galaxies. In fact, the extra terms 
appearing in the field equations of the $f(\mathcal{R},T)$ gravity theory may 
be treated as an extra fluid, whose source is due to the coupling between the 
matter and the geometry. 

Assuming the first fluid (the physical matter fluid) in the two fluid 
structure of $f(\mathcal{R},T)$ gravity theory as a dust-like perfect fluid, 
the modified Einstein equations have been solved and subsequently obtained the 
metric coefficients for a static spherically symmetric spacetime in the 
vicinity of GR. We have fixed the metric coefficients in terms of model 
parameters and derived tangential velocity relation for the test particle 
moving around the galaxy. The rotation curves generated by the relation 
\eqref{eq53} are nicely comparable to the observational data and they 
constrain $\delta$ to the order of $10^{- 6}$ -- $10^{- 5}$. After 
constraining the model parameters through fitting of theoretical and observed 
rotational curves of a sample of nineteen galaxies (10 LSB and 9 HSB), the 
best fit range of parameter $\beta$, that determines the strength of matter is 
found as $0.63<\beta<0.97$ and the product of $\lambda\beta$ is $<2$. 
Similarly, the parameter $a$ which determines the curvature coupling is found
to lie within the values of $700$ -- $800$ and the matter coupling 
parameter $b$ is found to remain within the range $0.2$ -- $5.6$. With 
these values of parameters our $f(\mathcal{R},T)$ model \eqref{eq35} explains 
the behavior of rotation curves of the selected sample of galaxies well. 
Further, in conformity to be with the observed data on the galactic rotation 
curves this result suggests that the parameter $\delta$ must be a considerably 
smaller one (of the order of $10^{-6}$ -- $10^{- 5}$) by confirming our 
assumption that $X(r)$ should slightly be different from the unity.  

Indeed, our specified model results in more or less the constant tangential 
velocity of a test particle needed for fitting observed rotation curves far 
away from the galactic center. The flatness behavior of rotation curves 
suggests a considerable amount of hidden matter in galactic halo and in other 
way, the $f(\mathcal{R},T)$ model we have considered leads to a desired result 
about the rotation curve through the modification of gravity. It therefore 
nullifies the presence of DM within the halo. In this study, no attempt has 
been made to include the Tully-Fisher relation in our work. We plan to do this 
in future with an attempt to extend the work choosing a non-minimally coupled 
$f(\mathcal{R},T)$ model by constraining the model parameters with the 
available observational data.

\section*{Acknowledgments}
UDG is thankful to the Inter-University Centre for Astronomy and Astrophysics
(IUCAA), Pune, India for the Visiting Associateship of the institute.


\begin{thebibliography}{99}
	\bibitem{Mar} M.~S.~Seigar, \emph{Dark Matter in the Universe},
	\href{https://doi.org/10.1088/978-1-6817-4118-5}{Morgan \& Claypool 
Publishers, USA (2015)}. 
	
	\bibitem{J_3} J.~H.~Oort, \emph{The force exerted by the stellar system in the direction perpindicular to the galactic plane and some related problems}, \href{https://adsabs.harvard.edu/full/1932BAN.....6..249O}{ Bulletin of the Astro. Ins. of the Netherlands, \textbf{6}, 249 (1932)}.		
	
	\bibitem{F_5a}  F. Zwicky, \emph{On the masses of Nabulae and of Clusters of Nabulae}, \href{https://adsabs.harvard.edu/full/1937ApJ....86..217Z} {Astrophys. J. \textbf{86}, 217 (1937)}.	
	
	\bibitem{Gar_4} K.~Garrett  and G.~Duda, \emph{Dark Matter: A Primer}, \href{https://www.hindawi.com/journals/aa/2011/968283/}{Adv.~Astron.~\textbf{2011}, 968283 (2011)} [\href{https://arxiv.org/abs/1006.2483}{arXiv:1006.2483}].	
	
	\bibitem{F_1}  F.~Zwicky, \emph{How Far Do Cosmic Rays Travel?}, \href{https://doi.org/10.1103/PhysRev.43.147}{Phys. Rev, \textbf{43}, 147 (1933)}. 	
	
	\bibitem{F_5} F. Zwicky, \emph{Republication of: The redshift of extragalactic nebulae}, \href{https://doi.org/10.1007/s10714-008-0707-4}{Gen.\ Rel.\ Grav.\  \textbf{41}, 207 (2009)}. 	
	
	\bibitem{H_9} H.\ W.\ Babcock, \emph{The rotation of the Andromeda Nebula}, \href{https://ui.adsabs.harvard.edu/abs/1939LicOB..19...41B/abstract}{Lick Observatory bulletins \textbf{498}, 41 (1939)}.		
	
	\bibitem{M_10}  M. S. Roberts, \emph{A High-Resolution 21-CM Hydrogen-Line Survey of the Andromeda Nebula}, \href{https://adsabs.harvard.edu/full/1966ApJ...144..639R}{ Astrophys. J. \textbf{144}, 639 (1966)}.
	
	\bibitem{K_12}  K. C. Freeman, \emph{On the Disks of Spiral and S0 Galaxies}, \href{https://adsabs.harvard.edu/full/1970ApJ...160..811F}{ Astrophys. J. \textbf{160}, 811 (1970)}.	
	
	\bibitem{J_8} J. C. Jackson, \emph{The Dynamics of Clusters of Galaxies in universes with non-zero cosmological constant, and the virial theorem mass discrepancies}, \href{https://doi.org/10.1093/mnras/148.3.249}{MNRAS  \textbf{148}, 249  (1970)}.	
	
	\bibitem{M_10a}  M. S. Roberts and  A. H. Rots, \emph{ Comparison of Rotation Curves of Different Galaxy Types}, \href{https://adsabs.harvard.edu/full/1973A%26A....26..483R}{Astron. and Astrophys. \textbf{26},  483  (1973)}. 	
	
	\bibitem{V_11} V. C. Rubin, and W. K. Ford, \emph{Rotation of the Andromeda Nebula from a Spectroscopic Survey of Emission Regions}, \href{https://doi.org/10.1086/150317}{Astrophys. J. \textbf{159}, 379 (1970)}.	
	
	\bibitem{V_11a} V. C. Rubin, W. K. Ford Jr. and N.~Thonnard, \emph{Extended Rotation Curves of High-Luminosity Spiral Galaxies. IV. Systematic 
Dynamical Properties, Sa$\rightarrow$Sc}, \href{https://adsabs.harvard.edu/full/1978ApJ...225L.107R}{ Astrophys. J. \textbf{225}, L107 (1978)}.	
	
	\bibitem{R_D}  D.\ H. Rogstad, and G.\ S.\ Shostak, \emph{Gross Properties of Five Scd Galaxies as Determined from 21-Centimeter Observations}, \href{https://adsabs.harvard.edu/full/1972ApJ...176..315R} { Astrophys. J. \textbf{176}, 315 (1972)}. 	
	
	
	\bibitem{L_2} L. \'{A}. Gegurgel  et al., \emph{Galactic rotation curves in brane world models}, \href{https://doi.org/10.1111/j.1365-2966.2011.18941.x}{MNRAS  \textbf{415} 3275, (2011)}  [\href{https://doi.org/10.48550/arXiv.1105.0159}{arXiv:1105.0159}].	
	
	\bibitem{Anna} A. Borriello and P. Salucci, \emph{The dark Matter Distribution in Disc Galaxies}, \href{https://academic.oup.com/mnras/article/323/2/285/1242888}{MNRAS  \textbf{323}, 285 (2001)} [\href{https://doi.org/10.48550/arXiv.astro-ph/0001082}{arXiv:astro-ph/0001082}].	

\bibitem{C15} C.~G.~B\"{o}hmer, T. Harko, F. S. N. Lobo, 
	\emph{Dark matter as a geometric effect in f(R) gravity},  \href{https://doi.org/10.1016/j.astropartphys.2008.04.003}{Astropart. Phys. \textbf{29}, 386 (2008)} [\href{https://doi.org/10.48550/arXiv.0709.0046}{arXiv:0709.0046}].	
	
	\bibitem{M_5b} M.~Heydari-Fard and H.~R.~Sepangi, \emph{Can local bulk effects explain the galactic dark matter?},
	\href{https://doi.org/10.1088/1475-7516/2008/08/018}{ JCAP \textbf{08}  018 (2008)} [\href{http://arxiv.org/abs/0808.2335v1}{arxiv.org/abs/0808.2335}].	
	
	\bibitem{Ca} C.~A.~Metzler et al., \emph{Weak Gravitational Lensing and Cluster Mass Estimates}, \href{https://iopscience.iop.org/article/10.1086/312144/pdf} {Astrophys. J. \textbf{520}, L9 (1999)}.	
	
	\bibitem{THar} T.~Harko, \emph{Galactic rotation curves in modified gravity with non-minimal coupling between matter and geometry},
	\href{https://doi.org/10.1103/PhysRevD.81.084050}{Phys. Rev. D  \textbf{81}, 084050 (2010)} [\href{https://doi.org/10.48550/arXiv.1004.0576}{arXiv:1004.0576}]	
	
	\bibitem{Tg}  T.~Matos,\, F. S. Guzm\'{a}n and D.~N\'{u}\~{n}ez, \emph{Spherical Scalar Field Halo in Galaxies}, \href{https://doi.org/10.1103/PhysRevD.62.061301}{ Phys. Rev. D \textbf{62}, 061301  (2000)}\ [\href{https://doi.org/10.48550/arXiv.astro-ph/0003398}{arXiv:astro-ph/0003398v2}].	
	
	\bibitem{V_11b} V.~C.~Rubin, \emph{Dark Matter in Spiral Galaxies},   
	\href{https://doi.org/10.1038/scientificamerican0683-96}{Scientific American \textbf{248}, 96 (1983)}.	
	
	\bibitem{V_11c} V.~C.~Rubin, \emph{The Rotation of Spiral Galaxies}, \href{https://doi.org/10.1126/science.220.4604.1339}{Science \textbf{220}, 4604  (1983)}.	
	
	\bibitem{Mil} M.~Milgrom, \emph{A modification of the Newtonian dynamics as a possible alternative to the hidden mass hypothesis}, \href{https://adsabs.harvard.edu/full/1983ApJ...270..365M}{Astrophys. J. \textbf{270}, 365 (1983)}.	
	
	\bibitem{Ad} A.~Dar, \emph{Tests of general relativity and Newtonian gravity at large distances and the dark matter problem}, \href{https://doi.org/10.1016/0920-5632(92)90192-U}{Nucl. Phys. B - Proc. Supl. \textbf{28}, 321 (1992)} 
[\href{https://doi.org/10.48550/arXiv.astro-ph/9407072}{arXiv:astro-ph/9407072}].	
	
	\bibitem{RH} R.~H.~Sanders, \emph{A historical perspective on Modified Newtonian Dynamics}, \href{https://doi.org/10.1139/cjp-2014-0206}{Can. J. Phys. \textbf{93}, 1 (2015)} [\href{https://doi.org/10.48550/arXiv.1404.0531}{arXiv:1404.0531}].	
	
	\bibitem{SC} S.~Capozziello and M.~De~Laurentis,\,  \emph{Extended Theories of Gravity}, \href{https://doi.org/10.1016/j.physrep.2011.09.003} {Phys. Rep. \textbf{509}, 167 (2011)} [\href{https://doi.org/10.48550/arXiv.1108.6266} {arXiv:1108.6266}].	
	
	\bibitem{HB} H. A. Buchdahl, \emph{Non-Linear Lagrangians and Cosmological Theory}, \href{https://doi.org/10.1093/mnras/150.1.1}{MNRAS  \textbf{150}, 
1 (1970)}.	
	
	\bibitem{Bra} A.~S.~Sefiedger, Z.~Haghani and H.~R.~Sepangi, \emph{Brane-f(R) gravity and dark matter},
	\href{https://doi.org/10.1103/PhysRevD.85.064012}{ Phys. Rev. D \textbf{85}, 064012  (2012)}\ [\href{https://arxiv.org/pdf/1202.4825.pdf}{arXiv:1202.4825}].	
	
	\bibitem{Ra6} R.~Zaregonbadi, M. Farhoudi and N. Riazi, \emph{Dark matter from f(R,T) gravity},  \href{https://doi.org/10.1103/PhysRevD.94.084052}{Phys. Rev. D   \textbf{94},  084052  (2016)} [\href{https://arxiv.org/abs/1608.00469}{arXiv:1608.00469}].
	
	\bibitem{Cap} S.~Capozziello, V. F. Cardone, and A. Troisi, \emph{Dark energy and dark matter as curvature effects},
	\href{https://doi.org/10.1088/1475-7516/2006/08/001}{ JCAP \textbf{0608}, 001, (2006)}  [\href{https://doi.org/10.48550/arXiv.astro-ph/0602349}{arXiv:astro-ph/0602349}]. 	
	
	\bibitem{Y11} Y.~Sobouti, \emph{An f(R) gravitation for galactic environments}, \href{https://doi.org/10.1051/0004-6361:20065188}{ Astron. and Astrophys.  \textbf{464}, 921 (2007)}  [\href{https://doi.org/10.48550/arXiv.astro-ph/0603302}{arXiv:astro-ph/0603302}]. 	
	
	\bibitem{cf} C.~F.~Martins and P.~Salucci, \emph{Analysis of Rotation Curves in the framework of $R^n$ gravity}, 
	\href{https://doi.org/10.1111/j.1365-2966.2007.12273.x}{MNRAS \textbf{381}, 1103 (2007)} [\href{https://doi.org/10.48550/arXiv.astro-ph/0703243}{arXiv:astro-ph/0703243}].
	
	\bibitem{SSh} S.~Shahidi and H. R. Sepangi, \emph{Brane worlds and dark matter}, \href{https://doi.org/10.1142/S0218271811018627} {Int. J. Mod. Phys. D \textbf{20}, 77 (2011)} [\href{https://doi.org/10.48550/arXiv.1009.5458}{arXiv:1009.5458}].
	
	\bibitem{MkM} M.~K.~Mak and T.~Harko, \emph{Can the galactic rotation curves be explained in brane world models?},	 
	\href{https://doi.org/10.1103/PhysRevD.70.024010}{Phys. Rev. D \textbf{70}, 024010 (2004)} [\href{https://doi.org/10.48550/arXiv.gr-qc/0404104}{arXiv:gr-qc/0404104}].	
	
	\bibitem{Ugu} U.~Camci, \emph{On Dark matter as a geometric effect in the galactic halo},
	\href{https://doi.org/10.1007/s10509-021-03997-5}{ Astrophys. Space. Sci. \textbf{366}, 91 (2021)}  [\href{https://doi.org/10.48550/arXiv.2109.09466}{arXiv:2109.09466}].	
	
	\bibitem{np} N. Parbin and U. D. Goswami, \emph{Galactic rotation dynamics in a new f(R) gravity model}, \href{https://doi.org/10.1140/epjc/s10052-023-11568-x} {Eur. Phys. J. C \textbf{83}, 411 (2023)}
	[\href{https://doi.org/10.48550/arXiv.2208.06564}{arXiv:2208.06564}].
	
	\bibitem{jr} J. R. Brownstein and J. W. Moffat, \emph{Galaxy rotation curves without nonbaryonic dark matter}, \href{https://doi.org/10.1086/498208}{Astrophys. J. \textbf{615}, 636 (2006)} [\href{https://arxiv.org/abs/astro-ph/0506370}{arXiv:astro-ph/0506370}]. 	
	
	\bibitem{bk} B.~K.~Yadav and M.~M.~Verma, \emph{Dark matter as scalaron in f(R) gravity models},
	\href{https://doi.org/10.1088/1475-7516/2019/10/052} {JCAP \textbf{10} 052 (2019)} [\href{https://arxiv.org/abs/1811.03964}{arXiv:1811.03964}].	
	
	\bibitem{JF} J.~F.~Jesu, S.~H.~Pereira, J.~L.~G.~Malatras and F.~A.~Oliveira, \emph{Can Dark Matter be a Scalar Field ?},
	\href{https://doi.org/10.1088/1475-7516/2016/08/046}{JCAP \textbf{08}, 046 (2016)} [\href{https://arxiv.org/abs/1504.04037}{arXiv:1504.04037}].	
	
	\bibitem{TT} T.~Tenkanen, \emph{Dark Matter from Scalar Field Fluctuations}, 
	\href{https://doi.org/10.1103/PhysRevLett.123.061302}{Phys. Rev. Lett. \textbf{123}, 061302 (2019)} [\href{https://doi.org/10.48550/arXiv.1905.01214}{arXiv:1905.01214}].	
	
	\bibitem{Sca} N. Parbin and U. D. Goswami, \emph{Scalarons mimicking Dark Matter in the Hu-Sawicki model of~f(R) gravity},
	\href{https://doi.org/10.1142/S0217732321502655}{Mod. Phys. Lets. A  \textbf{36}, 2150265 (2021)} [\href{https://doi.org/10.48550/arXiv.2007.07480}{arXiv:2007.07480}].	
	
	\bibitem{TH} T. Harko, F. S. N. Lobo, S. Nojri, S. D. Odintsov,\, \emph{f(R,T) gravity}, \href{https://doi.org/10.1103/PhysRevD.84.024020}{ Phys. Rev. D  \textbf{84}, 024020 (2011)} [\href{https://doi.org/10.48550/arXiv.1104.2669}{arXiv:1104.2669}].	
	
	\bibitem{pm} P. H. R. S. Moraes, \emph{The trace of the trace of the energy-momentum tensor-dependent Einstein’s field equations}, \href{https://doi.org/10.1140/epjc/s10052-019-7195-4}{Eur. Phys. J. C  \textbf{79}, 674 (2019)} [\href{https://doi.org/10.48550/arXiv.1907.04625}{arXiv:1907.04625}].	
	
	\bibitem{S7} S. Chakraborty, \emph{An alternative f(R,T) gravity theory and the dark energy problem}, \href{https://doi.org/10.1007/s10714-013-1577-y}{Gen. Rel. Grav. \textbf{45}, 2052 (2013)} [\href{https://doi.org/10.48550/arXiv.1212.3050}{arXiv:1212.3050}].	
	
	\bibitem{MJ} M. Jamil et al., \emph{Reconstruction of some cosmological models in f(R,T) gravity}, \href{https://doi.org/10.1140/epjc/s10052-012-1999-9}{ Eur. Phys. J. C \textbf{72}, 1999 (2012)} [\href{https://doi.org/10.48550/arXiv.1107.5807} {arXiv:1107.5807}].
	
	\bibitem{Hs} H. Shabani and M. Farhoudi, \emph{Cosmological and Solar System Consequences of f(R,T) Gravity Models}, \href{https://doi.org/10.1103/PhysRevD.90.044031}{Phys. Rev. D \textbf{90}, 044031 (2014)} [\href{https://doi.org/10.48550/arXiv.1407.6187}{arXiv:1407.6187}].	
	
	\bibitem{Eb} E. H. Baffou et al., \emph{Cosmological Evolution in f(R,T) theory with Collisional Matter}, \href{https://doi.org/10.1103/PhysRevD.92.084043}{Phys. Rev. D \textbf{92}, 084043 (2015)} [\href{https://doi.org/10.48550/arXiv.1504.05496}{arXiv:1504.05496}].
	
	\bibitem{Mjs}  M.~J.~S.~Houndjo,  \emph{Reconstruction of f(R,T) gravity describing matter dominated and accelerated phases}, \href{https://doi.org/10.1142/S0218271812500034}{Int. J. Mod Phys. D \textbf{21}, 1250003 (2012)} [\href{
https://doi.org/10.48550/arXiv.1107.3887}{arXiv:1107.3887}].	
	
	\bibitem{S.K.}  S.~K.~Maurya et al., \emph{Study on anisotropic strange stars in f(R,T) gravity: An embedding approach under simplest linear functional of matter-geometry coupling},
	\href{https://doi.org/10.1103/PhysRevD.100.044014}{Phys. Rev. D \textbf{100}, 044014 (2019)} [\href{https://arxiv.org/pdf/1907.10149.pdf}{arXiv:1907.10149}].	
	
	\bibitem{Jmz}  J.\ M.\ Z.\ Pretel, S.\ E.\ Joras, R.\ R.\ R.\ Reis and 
J.\ D.\ V.\ Arba\~{n}il, \emph{Neutron stars in f(R, T) gravity with conserved 
energy-momentum tensor: Hydrostatic equilibrium and asteroseismology}, 
	\href{https://doi.org/10.1088/1475-7516/2021/08/055}{JCAP \textbf{08}, 055 (2021)} [\href{https://doi.org/10.1088/1475-7516/2021/08/055}{arXiv:2105.07573}]	
	
	\bibitem{Ms} M. Zubair, S.  Waheed and Y. Ahmad, \emph{Static spherically symmetric wormholes in f(R,T) gravity}, \href{https://doi.org/10.1140/epjc/s10052-016-4288-1}{Eur. Phys. J. C  \textbf{76}, 444 (2016)} [\href{https://doi.org/10.48550/arXiv.1607.05998}{arXiv:1607.05998}].
	
	\bibitem{AVZ} A.~V.~Zasov, A.~S.~Saburova, A.~V.~Khoperskov, S.~A.~Khoperskov,\; \emph{Dark matter in galaxies},
	\href{https://doi.org/10.3367/UFNe.2016.03.037751}{Physics-Uspekhi, \textbf{60}, 3 (2017)} [\href{https://doi.org/10.48550/arXiv.1607.05998}{arXiv:1710.10630}].	
	
	\bibitem{Ps} P.~Salucci,\, \emph{The distribution of dark matter in galaxies},
	\href{https://doi.org/10.1007/s00159-018-0113-1}{Astron. \& Astrophys. Rev. \textbf{27}, 2  (2019)} [\href{https://arxiv.org/abs/1811.08843}{ 	arXiv:1811.08843}].	
	
	\bibitem{MPS} M.~Persic,\,P.~Salucci and F.~Stee,\, \emph{The universal rotation curve of spiral galaxies: I. The dark matter connection}, \href{https://doi.org/10.1093/mnras/278.1.27}{MNRAS  \textbf{281}, 27 (1996)} [\href{https://arxiv.org/abs/astro-ph/9506004}{arXiv:astro-ph/9506004}].	
	
	\bibitem{Bl} Bing-Lin Young,  \emph{A survey of dark matter and related topics in cosmology}, 
	\href{https://doi.org/10.1007/s11467-016-0583-4}{Front. Phys. \textbf{12}, 121201 (2017)}.
	
	\bibitem{TKO} T. Koivisto, \emph{Viable Palatini-f(R) cosmologies with generalized dark matter},
	\href{https://doi.org/10.1103/PhysRevD.76.043527} {Phys. Rev. D \textbf{76}, 043527 (2007)} [\href{https://doi.org/10.48550/arXiv.0706.0974}{arXiv:0706.0974}].
	
	\bibitem{H13} H.\ Shabani and M.\ Farhoudi,  \emph{f(R,T) Cosmological Models in Phase Space},
	\href{https://doi.org/10.1103/PhysRevD.88.044048}{ Phys. Rev. D  \textbf{88}, 044048  (2013)} [\href{https://arxiv.org/abs/1306.3164}{arXiv:1306.3164}].
	
	\bibitem{H3} H.\ Shabani and P.\ H.\ R.\ S. Moraes, \emph{Galaxy rotation curves in the f(R,T) gravity formalism},	\href{https://doi.org/10.1088/1402-4896/acd36d}{ Phys.\ Scr.\ \textbf{98}, 065302 (2023)} [\href{https://doi.org/10.48550/arXiv.2206.14920
}{arXiv:2206.14920}].
	
	\bibitem{FG} F.~G.~Alvarenga, M.~J.~S.~Houndjo, A.~V.~Monwanou and J. B.~Chabi Orou, \emph{Testing some f(R,T) gravity models from energy conditions}, \href{https://doi.org/10.4236/jmp.2013.41019} {J. Mod. Phys. \textbf{4}, 130 (2013)} [\href{https://doi.org/10.48550/arXiv.1205.4678}{arXiv:1205.4678}].
	
	\bibitem{M3} M.~Sharif and M.~Zubair,  \emph{Anisotropic Universe Models with Perfect Fluid and Scalar Field in f(R,T) Gravity}, \href{https://doi.org/10.1143/JPSJ.81.114005} {J. Phys. Soc. Jpn. \textbf{81}, (2012)} [\href{https://arxiv.org/abs/1301.2251} {arXiv:1301.2251}].	
	
	\bibitem{JB} J.~Barrientos O. and G.~F. Rubilar,  \emph{Comment on ``f(R,T) gravity"}, \href{https://doi.org/10.1103/PhysRevD.90.028501} {Phys. Rev. D \textbf{90}, 028501 (2014)}. 	
	
	\bibitem{Fra} F. S. N. Lobo and T. Harko, \emph{Extensions of f(R) Gravity : Curvature-Matter Couplings and Hybrid Metric-Palatini Theory}, \href{https://doi.org/10.1017/9781108645683}{Cambridge University Press, (2018)}.	
	
	\bibitem{R5} R.~Zaregonbadi and M.~Farhoudi, \emph{Cosmic Acceleration From Matter–Curvature Coupling},
	\href{https://doi.org/10.1007/s10714-016-2137-z}{Gen. Rel. Grav.  \textbf{48}, 142 (2016)} [\href{https://arxiv.org/abs/1512.05604}{arXiv:1512.05604}].
	
	\bibitem{TK1} T.~Koivisto, \emph{A note on covariant conservation of energy–momentum in modified gravities}, \href{https://doi.org/10.1088/0264-9381/23/12/N01}{Class. Quant. Grav. \textbf{23}, 4289 (2006).} 	
	
	\bibitem{J9} J.~Binney and S.~Tremaine, \emph{Galactic Dynamics},  \href{https://press.princeton.edu/books/ebook/9781400828722/galactic-dynamics}{Princeton University Press, Princeton (1987)}.	
	
	
	\bibitem{M14} M.~P.~Hobson,\, G.~P.~Efstathiou\, and\, A.~N.~Lasenby, \emph{General Relativity-An Introduction for Physicists}, 
	\href{https://doi.org/10.1017/CBO9780511790904}{Cambridge University Press, New York, pp \textbf{205} (2006).}
	
	\bibitem{rj} R. J. A. Lambourne, \emph{Relativity, Gravitation and 
Cosmology}, Cambridge University Press, pp \textbf{139} (2010).
	
	\bibitem{af} A. Finch and J. L. Said, \emph{Galactic rotation dynamics in $f(T)$ gravity}, \href{https://doi.org/10.1140/epjc/s10052-018-6028-1}{Eur. Phys. J. C  \textbf{78}, 560 (2018)} [\href{https://doi.org/10.48550/arXiv.1806.09677}{arXiv:1806.09677}].
	\bibitem{wa} R. Wald, \emph{General Relativity}, The University of Chicago Press, Chicago and London, pp \textbf{77} (1984).
	
	\bibitem{spark}	F. Lelli, S. S. McGaugh and J. M. Schombert, \emph{SPARC: Mass Models for 175 Disk Galaxies with Spitzer Photometry and Accurate Rotation Curves}, \href{https://doi.org/10.3847/0004-6256/152/6/157}{Astron. J. \textbf{152}, 157 (2016)}[\href{https://arxiv.org/abs/1606.09251}{arXiv:1606.09251}]; \href{http://astroweb.cwru.edu/SPARC/MassModels_Lelli2016c.mrt}{http://adsabs.harvard.edu/abs/2016arXiv160609251L}.
	\bibitem{pd} P. D. Mannheim and J. G. O’Brien, \emph{Fitting galactic rotation curves with conformal gravity and a global quadratic potential}, \href{https://doi.org/10.1103/PhysRevD.85.124020}{Phys. Rev. D \textbf{85}, 124020 (2012)} [\href{https://doi.org/10.48550/arXiv.1011.3495}{arXiv:1011.3495}].
	
	\bibitem{PL} P. Li et.al. \emph{Fitting the radial acceleration relation to individual SPARC galaxies}, \href{https://doi.org/10.1051/0004-6361/201732547}{ Astron. and Astrophys. \textbf{615}, A3 (2018)} [\href{https://arxiv.org/abs/1803.00022}{arXiv:1803.00022}].
	
\end{thebibliography}
\end{document}